\documentstyle[twocolumn,aps]{revtex}

\input{psfig}

\begin{document}

\draft

\title{The Liquid Scintillator Neutrino Detector and LAMPF Neutrino Source}

\author{C. Athanassopoulos$^{12}$, L. B. Auerbach$^{12}$,
 D. Bauer$^3$,
R. D. Bolton$^7$, R. L. Burman$^7$,\\
I. Cohen$^6$, D. O. Caldwell$^3$, B. D. Dieterle$^{10}$, J. B. Donahue$^7$,
A. M. Eisner$^4$,\\ A. Fazely$^{11}$,
F. J. Federspiel$^7$, G. T. Garvey$^7$, M. Gray$^3$, R. M. Gunasingha$^8$,
V. Highland$^{12,13}$,\\ R. Imlay$^8$, K. Johnston$^{9}$,
 H. J. Kim$^8$, W. C. Louis$^7$, A. Lu$^3$,
J. Margulies$^{12}$,  G. B. Mills$^7$, K. McIlhany$^{1}$,\\ W. Metcalf$^8$,
R. A. Reeder$^{10}$, V. Sandberg$^7$,
M. Schillaci$^7$, D. Smith$^5$,\\I. Stancu$^{1}$, W. Strossman$^{1}$,
R. Tayloe$^{7}$, G. J. VanDalen$^{1}$,
W. Vernon$^{2,4}$, Y-X. Wang$^4$,\\ D. H. White$^7$, D. Whitehouse$^7$,
D. Works$^{12}$, Y. Xiao$^{12}$,
S. Yellin$^3$}
\address{$^1$University of California, Riverside, CA 92521}
\address{$^2$University of California, San Diego, CA 92093}
\address{$^3$University of California, Santa Barbara, CA 93106}
\address{$^4$University of California
Intercampus Institute for Research at Particle Accelerators,
Stanford, CA 94309}
\address{$^5$Embry Riddle Aeronautical University, Prescott, AZ 86301}
\address{$^6$Linfield College, McMinnville, OR 97128}
\address{$^7$Los Alamos National Laboratory, Los Alamos, NM 87545}
\address{$^8$Louisiana State University, Baton Rouge, LA 70803}
\address{$^9$Louisiana Tech University, Ruston, LA 71272}
\address{$^{10}$University of New Mexico, Albuquerque, NM 87131}
\address{$^{11}$Southern University, Baton Rouge, LA 70813}
\address{$^{12}$Temple University, Philadelphia, PA 19122}
\address{$^{13}$Deceased}

\date{\today}
\maketitle
\begin{abstract}
A search for neutrino oscillations of the type $\bar\nu_{\mu}
\to \bar\nu_{e}$ has been conducted at the Los Alamos Meson Physics
Facility using $\bar\nu_{\mu}$ from muon decay at rest.
Evidence for this transition has been reported previously.
This paper discusses in detail the experimental setup, detector operation
 and neutrino source, including aspects relevant to oscillation searches
in the muon decay-at-rest and pion decay in flight channels.
\end{abstract}

\section*{1.\ \ Introduction}%

\subsection*{1.1\ \ Motivation}%

Standard electroweak theory has three lepton families, each with a charged and
neutral partner.
It is widely accepted that lepton family number is conserved and so these
families are not expected to transform into one another.
The minimal theory also assumes that neutrino masses are zero, providing a
natural explanation for the left-handed character of the weak interaction.
In the last few years it has gradually become accepted that the standard model
may well need to be extended to accept finite neutrino mass.
The deficit of solar neutrinos and indications from anomalies in atmospheric
neutrino events have raised expectations that neutrino oscillations may occur,
implying lepton family number mixing and finite neutrino masses.
A search for neutrino oscillations has been conducted with the apparatus
described in this paper, both from neutrinos generated by muon decay at rest
as well as those from pion decay in flight.
The source is unique in that the backgrounds expected from
conventional processes are very small.

\subsection*{1.2\ \ Experimental Method}%

The Liquid Scintillator Neutrino Detector (LSND) was designed to detect
neutrinos originating in a proton target and beam stop at the Los Alamos
 Meson Physics Facility (LAMPF).
The primary goal was to search for transitions from muon-type to
electron-type neutrinos in two complementary ways.
The LSND experiment was proposed \cite{proposal} for this purpose in
1989 and first operated in 1993.
First results have been published \cite{paper1} on the search for the
appearance
 of $\bar\nu_{e}$ from the large flux of $\bar\nu_{\mu}$ from muon decay at
rest (DAR).
A search is also being conducted for electron neutrinos in a dominantly
${\nu_{\mu}}$ beam from pions that decay in flight (DIF) \cite{dif}.
This paper focuses on the apparatus that was used in both searches.
Details of the experiment are included that could not appear in a more
compressed format.

For the``at rest'' experimental strategy to be successful, the target had
to be a copious source of $\bar\nu_{\mu}$ in a particular energy range of
interest with relatively few $\bar\nu_{e}$ produced by conventional means.
The detector had to be able to recognize interactions of $\bar\nu_{e}$ with
precision and separate them from other neutrino types, including the
large rate of ${\nu_e}$ expected from the target.
The observation of $\bar\nu_{e}$ in excess of the number expected from
conventional sources may be interpreted as evidence for neutrino
oscillations.
However, although we will concentrate on the oscillation hypothesis,
it must be noted that any unconventional process that creates $\bar\nu_{e}$
either at production, in flight, or in detection could produce a
positive signal in this search.
Lepton number violation in muon decay
$\mu^{+}\to e^{+}+\bar\nu_{e}+{\nu_{\mu}}\ \ \ \ $ is a good example.

The accelerator and water target produced pions copiously.
Most of the positive pions came to rest, and decayed through the sequence
$$\pi^{+}\to\mu^{+}+\nu_{\mu}\ \ ,$$
$$\mu^{+}\to e^{+}+{\nu_e}+\bar\nu_{\mu}\ \,$$
supplying $\bar\nu_{\mu}$ with a maximum energy of 52.8 ${\,{\rm~MeV}}$.
The energy dependence of the $\bar\nu_{\mu}$ flux from decay at rest is
very well known, and the absolute value is known to 7\% \cite{burman}.
The open space around the target was short compared to the pion decay
length, so only a small fraction of the pions (3.4\%) decayed in flight
through the first reaction.
A much smaller fraction (approximately 1\%) of the muons decayed in flight,
due to the difference in lifetimes.

The chain starting with $\pi^-$ produced only a small number of $\bar\nu_{e}$,
because most negative pions and muons are absorbed.
In the LAMPF proton beam, and with a water target, positive pion production
 exceeded that of negative pions by a factor of about eight.
Negative pions which came to rest in the beam stop and shielding were
captured before decay occurred, so only the pions which decayed
 in flight contributed to a $\bar\nu_{e}$~background.
Virtually all of the negative muons that arose from pion decay in flight came
to rest in the beam stop before decaying.
Most were captured from atomic orbit, a process which yields
${\nu_{\mu}}$; the remaining 12\% decayed and produced $\bar\nu_{e}$.
The relative yield, compared to the positive channel, was estimated to be
$\sim (1/8) * 0.034 * 0.12 \approx 5 \times 10^{-4}$.
As is discussed below, a detailed simulation was used to predict neutrino
fluxes.

Charged current reactions in the detector were dominated by ${\nu_e}$ on
$^{12}C$.
Electrons from this reaction have energies below 36${\,{\rm~MeV}}$ because of
the mass
difference of $^{12}C$ and $^{12}N$.
LSND detected $\bar\nu_{e}$ through the reaction $${\bar\nu_{e} + p \to e^+ +
n\ \ \ ,}$$
a process with a well-known cross section \cite{llewellyn_smith},
followed by the neutron-capture reaction
$$n + p \to d + {\gamma}~(2.2{\,{\rm~MeV}})\ \ \ $$
The detection signature consisted of an electron-like signal, followed
by a $2.2{\,{\rm~MeV}}$ photon correlated with the first signal in both
position
and time.
Although it was not possible to distinguish
an $e^-$ from an $e^+$, reactions due to background ${\nu_e}$ could not produce
such a correlated photon for events with electron energy above
20${\,{\rm~MeV}}$.
This value was because of the energy required to eject a neutron in a charged
current reaction.
The requirement of an $e^\pm$ energy above 36${\,{\rm~MeV}}$
eliminated most of the ${\nu_e}$
background due to an accidental coincidence with a uncorrelated ${\gamma}$
signal.
For the decay in flight search electrons above $60 {\,{\rm~MeV}}$ are
identified
from ${\nu_e} ~^{12}C \rightarrow e^- X$ and $ \bar\nu_{e} ~^{12}C
\rightarrow e^+ X.$
The electron energy spectrum from DIF is expected to be broader than from DAR
but the background from conventional neutrino events is expected to be much
less.

\begin{figure}[ht]
\centerline{\psfig{figure=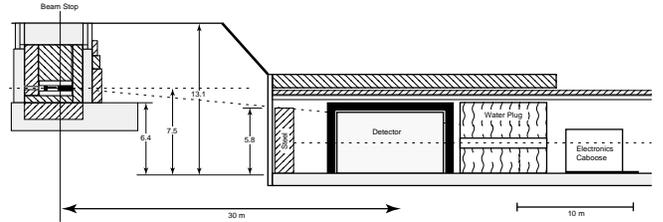,width=3.4in,silent=}}
\caption{Detector enclosure and target area configuration, elevation view}
\end{figure}

The detector was located about $30{\,{\rm~m}}$ from the neutrino source and was
shielded by the equivalent of $9{\,{\rm~m}}$ of steel.
A schematic of the layout is shown in Fig. 1.
The detector was under $\sim 2\,$kg/{\,{\rm cm}}$^2$ of overburden reducing the
cosmic
ray flux significantly from that at the surface.
A liquid scintillator veto shield surrounded the detector on all sides except
on the bottom.
The detector was a tank filled with 167 metric tons of mineral oil (CH$_{2}$),
with a small admixture (0.031 g/l) of butyl PBD scintillant.
This dilute mixture allowed the detection of both {$\check{\rm C}$erenkov }
light and isotropic
scintillation light.
This resulted in robust particle identification for $e^\pm$, location of the
event vertex in space, and a measurement of the $e^\pm$ direction.
The light was detected by 1220 8$^{\prime\prime}$ PMTs, covering $\sim 25\%$
of the surface inside the tank wall.
Each channel was digitized for pulse height and time.
The electronics and data acquisition systems were designed explicitly to detect
and correlate events separated in time.
This was necessary both for many neutrino induced reactions and for cosmic
ray backgrounds.
The behavior of the detector was calibrated using a large sample of ``Michel''
$e^\pm$ from the decays of stopped cosmic ray muons.
These $e^\pm$ were in just the right energy range for the
$\bar\nu_{\mu}\to\bar\nu_{e}$
search.

Even with this shielding, there remained a large background to the
oscillation search due to cosmic rays, which needed to be suppressed by
about nine orders of magnitude to reach a sensitivity limited by the
neutrino source.
The cosmic ray muon rate through the tank was $\sim 4${\,{\rm~kHz}}, of which
$\sim 10\%$ stopped and decayed in the scintillator.
Details of the suppression of this background in the DAR search will be
discussed in reference \cite{bigpaper2}.
Finally, any remaining cosmic ray background was very well measured because
about 13 to 14 times as much data were collected when the beam was off as on.
The result of these procedures was to reduce the cosmic ray background
below the level of sensitivity required for the decay at rest oscillation
search; similar techniques were found useful for the DIF oscillation search.

\subsection*{1.3\ \ Outline of this paper}%

	After the introduction, we present a description of the
neutrino source used in the LSND experiment in chapter two.
The detector hardware are described in chapter three and the data acquisition
is in chapter four.
Chapter five covers basic detector simulation.
Chapter six covers event reconstruction and calibration of the detector.

\section*{2.\ \ Neutrino Source}%

\subsection*{2.1\ \ LAMPF}%

\subsubsection*{2.1.1\ \ Accelerator description}%

The linear accelerator at Los Alamos used conventional ion sources for protons
and for H$^{-}$ ions.
Ion sources provided particles for acceleration that were selected on a
pulse-by-pulse basis.
Protons passed through a transition section to a drift tube linear accelerator
 of Alvarez type operating at 201.25 MHz.
Protons were accelerated in this section to 50 MeV and then injected directly
into a side-coupled linac structure operating at 805 MHz for acceleration to
a nominal energy of 800 MeV.
The acceleration process was repeated at 120 Hz.
During the course of this experiment, all cavities were operational in the
linac structure.
This ensured that the output proton beam kinetic energy of 800 MeV was constant
 to much better than 1\%.
The linac output beam was transported to a high intensity area shown in
Fig. 2.
The output beam energy has been measured in this configuration, and the
results are discussed in section 2.3.

\begin{figure}[ht]
\centerline{\psfig{figure=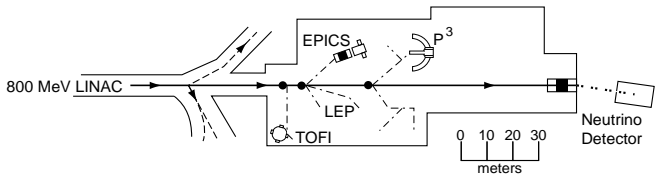,width=3.4in,silent=}}
\caption{Schematic of the high intensity experimental area}
\end{figure}

\subsubsection*{2.1.2\ \ Time structure of beam}%

The linear accelerator at times accelerated both proton and H$^{-}$ beams in
interleaved pulses.
During much of the data taking in the period of the experiment, the proton
beam was delivered at 120 Hz to the experimental area described below.
For part of the run in 1993 and 1995, H$^{-}$ beam at 20 Hz was delivered
elsewhere and those pulses were not available to this experiment.
Each of the 120-Hz proton pulses was approximately $600{~\mu s}$ long and had a
substructure that consisted of pulses approximately 0.25 ns wide at 201.25 MHz.

The beam current at the output of the accelerator was typically 1 mA.
The accelerating RF frequency was set for optimum beam acceleration in the
linac
 and remained constant to parts in $10^{7}$ for the entire data taking period.
The phase of the accelerated beam relative to the RF waveform was varied for
optimum accelerator performance.
The repetition rate was approximately synchronized to the line frequency so
that
 long-term variation of this rate occurred at the level of 0.1\% in a day.
A beam permit signal (later referred to as H$^{+}$) was generated preceeding
 acceleration and lasting through beam delivery.
The gate that was generated by this signal was used after the fact to verify
the
relative timing of neutrino events and the accelerator cycle.

The existence of this beam permit signal was necessary for beam on-off
subtraction.
The amplitude of the beam typically varied by less than 1\% over short periods,
with overall beam availability in excess of 90\%.

\subsubsection*{2.1.3\ \ Upstream beam use}%

This experiment was performed at LAMPF where a broad-based program of
nuclear science was carried out at the same time that the measurements of
this experiment were being made.
The program depended on secondary beams produced at two targets substantially
upstream of the neutrino target A6.
The two targets A1 and A2 were used to generate pion and muon beams for
experimentation.
Since pions were produced at these targets and subsequent decay occurred, they
were also potential sources of neutrinos in LSND.
Pions mostly decayed in the evacuated enclosures near these targets.
The targets A1 and A2 were made of carbon, 3 and 4 ${\,{\rm cm}}$ thick,
respectively.
An estimate of the relative flux with respect to A6 for DAR was made by simply
observing that the production rate was 25\% of that at the neutrino target A6
and the distance was typically four times further from the neutrino detector.
This gave a neutrino flux from this source of about 1.5\% that from A6.
The detailed calculation is discussed in sections 2.4 for DAR and 2.5 for DIF
below.

\subsection*{2.2\ \ Target geometry and shielding}%

Fig. 3 shows a plan view of the A6 target area and
Fig. 4 shows an elevation view.

\begin{figure}[htp]
\centerline{\psfig{figure=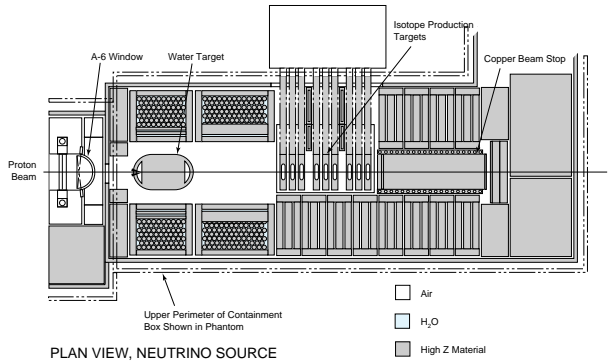,width=3.2in,silent=}}
\caption{Plan view of the target box at area A6}
\end{figure}

\begin{figure}[htp]
\centerline{\psfig{figure=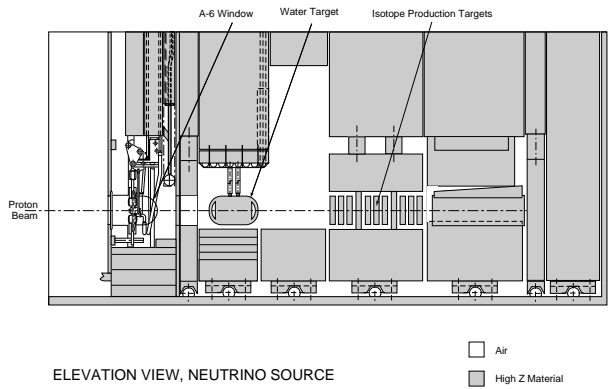,width=3.2in,silent=}}
\caption{Elevation view of the target box at area A6}
\end{figure}

The proton beam entered from the left, passed through the water target and
ended in the beam stop.
The target consisted of an outer inconel vessel filled with water and fitted
with
baffles to direct flow.
A larger scale diagram of the target used in '94/'95 is shown in Fig. 5.

\begin{figure}[ht]
\centerline{\psfig{figure=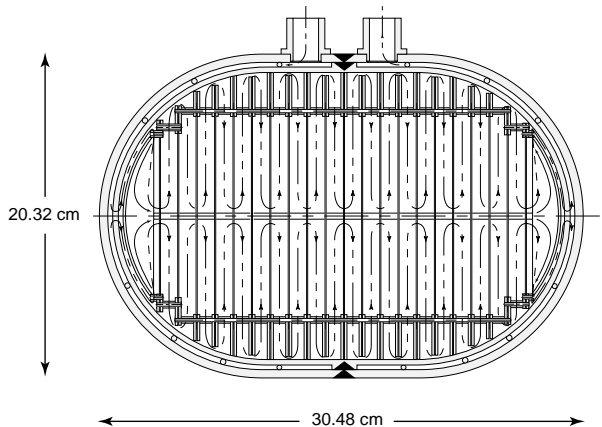,width=3.2in,silent=}}
\caption{Construction schematic of the water target assembly used in '94/'95}
\end{figure}

Rapid water flow, necessary to prevent boiling during beam passage, was
 monitored to verify target conditions.
The ratio of mass of inconel to water was about 25\% for the proton beam, which
passed normally through the baffles; this ratio was slightly larger for
secondary
 pions that were produced at finite angles.
Water-cooled iron shielded the entire region.
Within this shielding around and downstream of the target was a volume in which
pion decay occurred.
The shielding was hermetically sealed, partly because neutron background would
be a serious problem if the shielding were not complete and partly because
activation of the air near the target region would be a health hazard if the
air were not contained.
The DAR part of the experiment depended on suppression of the decay of negative
 muons, which was accomplished first by the fact that negative pions that
stop were all absorbed.
Those negative pions that decayed in flight before coming to rest
produced negative muons that almost entirely stopped either in iron, copper or
in water-cooling channels.
Typically, 9\% of $\mu^-$ stop in water, almost entirely in the production
target.
A typical pion had a momentum of 200 MeV/$c$ and $\beta {\gamma} c \tau$ =
12 m.
This was long compared to a typical trajectory and resulted in a fraction of
pions that decay in flight of about 3.4\%.
The inserts in Fig. 3 and 4 that are labelled
``isotope production targets'' were targets for producing radioisotopes and
were
responsible for a small amount of pion production, which was taken into account
 in the simulation of neutrino flux.

In Fig. 2 is shown two upstream targets A1 and A2, which were also
included in the simulation of neutrino flux.
Each of these targets was mounted in an enclosure which was surrounded closely
by hermetic shielding except for vacuum pipes which permitted egress for four
secondary beams, one on each side for both targets.

\subsection*{2.3\ \ Characteristics of 1993, 1994 and 1995 Runs}%

The beam conditions in 1993, 1994 and 1995 were slightly different.
In 1993 the target thickness at A1 was $3{\,{\rm cm}}$ and at A2 $2{\,{\rm
cm}}$, both carbon
targets of uniform thickness along the beam.
In 1994 and 1995, A2 was increased to $4{\,{\rm cm}}$.
The kinetic energy of protons from the linac was 797 MeV and was known to
about 1 MeV.
After the A1 target the beam emerged at 786.5 MeV and after A2 it was 767.6
MeV.
The proton current loss was 85 $\mu$A at A1 and 103 $\mu$A at A2.
These conditions were quite stable through the running period in 1994 and 1995.
In 1993 the A2 target suffered some erosion of thickness, which was
tracked and compensated.
The neutrino water target was $20{\,{\rm cm}}$ in thickness in 1993,
$30{\,{\rm cm}}$ in 1994 and 1995.
This increase in thickness, together with the slightly lower energy due to the
increase in A2 thickness in 1994, resulted in an increase in the DIF
${\nu_{\mu}}$ flux  of 8\%.
The target was not in place for part of the time in 1995, which was accounted
for
in the total flux calculations.
The charge delivered to the beam stop in 1993 was 1787 C, in 1994 it was 5904
C,
in 1995 4794 C with target in and 2286 C with target out.
Each of these charge accumulations was with the detector in operation and
taking
 data.

\subsection*{2.4\ \ Decay at Rest Neutrino Flux Calculation and Measurement}%

A detailed simulation was used to predict neutrino fluxes at the
detector and is described in \cite{burman}.
 Measurements of pion production fluxes that
were taken at a number of proton energies at LAMPF \cite{pion_fluxes}
were used as input to this simulation.
The detailed geometry of the sources (A1, A2, and A6) described above were
included in the simulation.
In addition, the yield of muon decays was determined in a separate experiment
 \cite{beam_mockup} in which a proton beam of variable energy was introduced
into a composite structure of water and copper to approximately duplicate the
materials in which pion production occurred in this experiment.
The overall number of muon decays per incident proton was determined in this
subsidiary experiment, and the neutrino flux calculation was renormalized
slightly to fit the measured yield in the separate experiment.
It should be noted that this procedure normalized the neutrino yield from
positive pions quite well, but since the negative pions were almost entirely
 absorbed, the measurement was insensitive to the consequences of the negative
 pion decay rate.
The shape of the neutrino flux from $\pi^+$ and $\mu^+$ DAR is well known
and is shown in Fig. 6, so that only the absolute amplitude was
determined from experiment and simulation.

\begin{figure}[ht]
\centerline{\psfig{figure=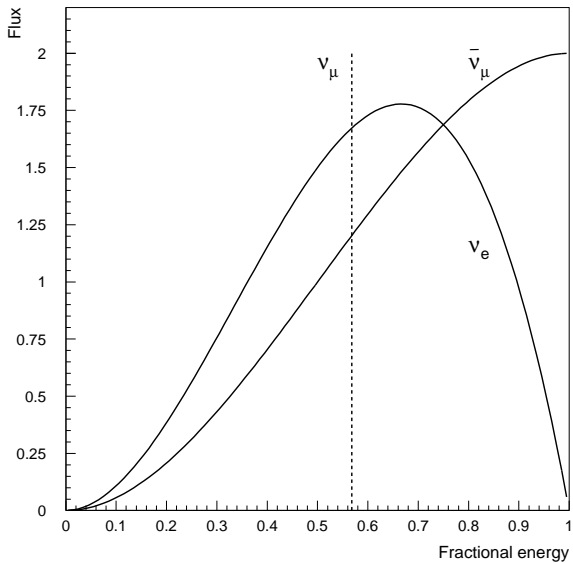,width=3.4in,silent=}}
\caption{Flux shape of neutrinos from pion and muon decay at rest}
\end{figure}

The $\bar\nu_{\mu}$ flux integrated over energy for the configuration
described here was $7.6\times 10^{-10}\ \bar\nu_{\mu}$ /{\,{\rm
cm}}$^{2}$/proton
averaged over the fiducial volume of the detector.
The ${\nu_e}$ flux is identical to the $\bar\nu_{\mu}$ flux.
The DAR simulation depended on the geometry at the three target
stations in which substantial beam interacted.
Beam loss in the accelerator and in locations other than target areas was
very small because operational and health considerations made it imperative
to lose very little of the proton beam in other than specially hardened areas.
The targets at A1 and A2 were surrounded by enclosures that were typically
$50{\,{\rm cm}}$ in dimension with closely packed steel shielding, apart from
relatively small apertures for beams to emerge.
The distance from A1 to A6 was 107 m and A2 to A6 was 82 m.
The calculation of the DAR flux was then relatively simple, and the
DAR spectra at the detector was governed by solid-angle
considerations and the interaction fraction in the target.
Pion DAR was isotropic and muon decay was effectively isotropic
because spin precession and the varying production angle washed out angular
effects.
There was one exception to this, which is discussed in the next section:
that part of pion production that occurred at small angles and thus was
able to pass down the beam pipe for a significant distance.
The ``in-flight'' decay probability was then much larger than in the target
enclosures, and subsequent muon decay probability in flight was also enhanced.
Detailed simulation gave a DAR flux at the detector from positive
pion and muon decay that was 0.5\% from A1, 1\% from A2, and 98.5\% from A6.

An additional corroboration exists from the experiment \cite{e225} performed at
90$^{\circ}$ to the same target that measured neutrino-electron elastic
scattering as a test of Standard Electroweak theory.
This cross section of ${\nu_e}$ scattering from electrons is now well known
and can be used to estimate the flux in contrast to the procedure of that
previous experiment.
This provided an independent confirmation of the flux estimate used here with
an uncertainty of 18\%.

\subsection*{2.5\ \  Pion Decay in Flight Flux}%

\subsubsection*{2.5.1\ \ ${\nu_{\mu}}$, $\bar\nu_{\mu}$ Flux calculation}%

The calculation of the neutrino flux (${\nu_{\mu}}$) from pion DIF
at the A6 beam stop, A1 and A2, proceeded in a similar way to the calculation
 of the DAR flux.
The same input data were used to calculate pion production normalized in the
same way as the data in the previous section.
The overall positive pion DAR flux was known to 7\% as constrained by the
simulation
experiment described previously.
The negative pion flux lacks this constraint.
However, the same normalization factor was assumed to apply
because it came mainly from the effects of thick target particle production
of $\pi^\pm$ by protons.
It was believed that the geometry of the decay volumes, including the
neighborhood of A1 and A2, is well known from the constraints attendant in
mechanical assembly.
A lack of knowledge of the shape of the pion production spectrum folded
into this geometry introduced some additional uncertainty.
These uncertainties have been estimated to increase the systematic error on the
DIF fluxes to 12\% in absolute magnitude \cite{burman_technote}.
A significantly smaller point-to-point uncertainty may be expected in the
spectrum.
The ${\nu_{\mu}}$ spectrum shown in Fig. 7 was
at the center of the detector for the three years described in this paper.
The integrated flux over the detector for this period was $6.2\times 10^{12}$
 ${\nu_{\mu}}$/{\,{\rm cm}}$^{2}$.
Although the flux at the center of the detector is shown in the figures,
the integrated flux used a calculation of flux throughout the tank.
As was mentioned in section 2.3 the water target was retracted from the proton
beam for part of the data taking in 1995.
With the water target out the DIF flux was $3.6\times 10^{-11}$
$\nu_{\mu}/{\,{\rm cm}}^{2}$/proton.

\begin{figure}[htp]
\centerline{\psfig{figure=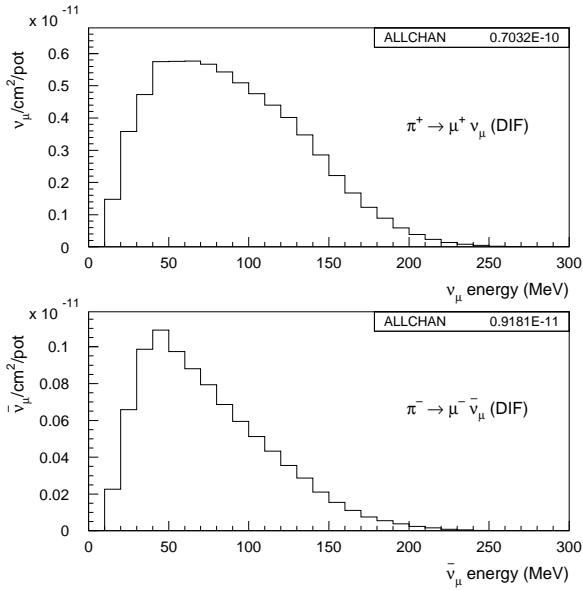,height=3.4in,silent=}}
\caption{Calculated DIF energy spectra for ${\nu_{\mu}}$ and $\bar\nu_{\mu}$ at
the center
of the detector}
\end{figure}

A similar calculation was performed for negative pions that decay in flight,
leading
to the $\bar\nu_{\mu}$ spectrum also shown in Fig. 7 and an
integrated flux of $8.0\times 10^{11} \bar\nu_{\mu}$/{\,{\rm cm}}$^{2}$
averaged
over the detector volume.

\subsubsection*{2.5.2\ \ ${\nu_e}$ and $\bar\nu_{e}$ flux calculation}%

Electron neutrinos were produced (occasionally) from pion decay in flight
followed by muon decay in flight.
The rare branching mode $\pi^{+}\to e^{+}+\nu_{e}$ also contributed to the
$\nu_{e}$ flux.
These contributions have been calculated and are shown in
Fig. 8.

\begin{figure}[htp]
\centerline{\psfig{figure=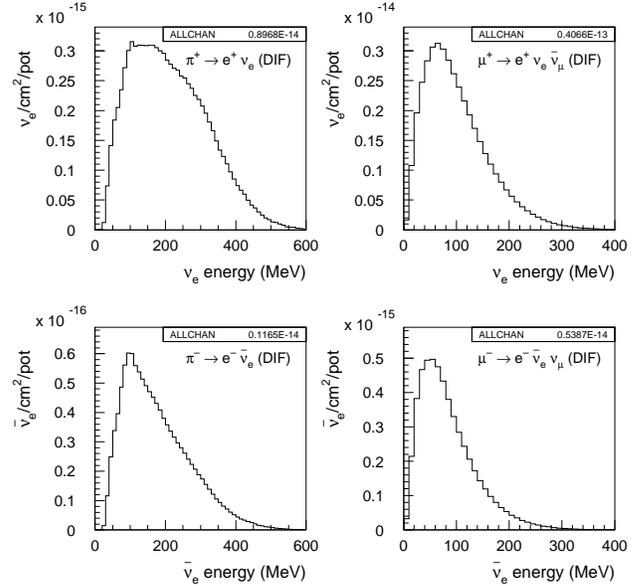,height=3.4in,silent=}}
\caption{The left two figures are ${\nu_e}$ and $\bar\nu_{e}$ flux at detector
center from pion decay.  The right two figures are similarly calculated
${\nu_e}$ and $\bar\nu_{e}$ fluxes from muon decay.}
\end{figure}

\subsection*{2.6\ \ Decay at Rest $\bar\nu_{e}$ Flux}%

Negative muons were produced through the chain of negative pion production
and decay in flight.
This was followed almost always by either decay or absorbtion of the negative
muon in the field of a nucleus.
This decay process in this experiment occurred dominantly in Fe and Cu although
a fraction (35\%) of decays were in oxygen in the water of the pion
production target and in the water cooling of the shield.
A $\mu^{-}$ in the lowest atomic state in Fe was contained in a
potential of $\sim 1.9$ {\,{\rm~MeV}}, which influenced the momentum
distribution of
the $\mu^{-}$.
In turn, this momentum distribution affected the decay spectrum of the outgoing
 leptons.
These effects have been calculated by a number of authors for the electron
\cite{capture_theory} but not for the neutrino spectrum.
The actual capture rates are well described by theory, and the data in
\cite{capture_expt} have been used to estimate the $\bar\nu_{e}$ rate expected
in
the experiment.
The fraction of produced $\mu^-$ that decay has been estimated at about 12\%.
The energy spectrum of $\bar\nu_{e}$ from this source is approximately
$x^2 (1 - x)$, where $x$ is the ratio of neutrino energy to the end
point energy of the spectrum (52.8${\,{\rm~MeV}}$).
This spectrum is significantly softer than that expected from $\bar\nu_{\mu}$
to
 $\bar\nu_{e}$ oscillations, $x^2 (3 - 2x)$.
The overall yield from bound muon decay was subject to the same systematic
error as the pion DIF flux.

\section*{3.\ \ Detector Hardware}%

\subsection*{3.1\ \ Detector, Veto, and Shielding}%

\subsubsection*{3.1.1\ \ Detector description}%

The experimental detector was situated in the enclosure shown in
Fig. 1.
The detector proper was contained in a steel tank roughly cylindrical in shape,
 8.3 m long internally, with a diameter of 5.7 m, as shown schematically in
Fig. 9.

\begin{figure}[ht]
\centerline{\psfig{figure=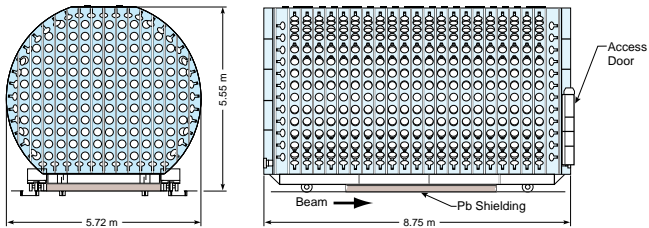,width=3.4in,silent=}}
\caption{Schematic of the detector tank}
\end{figure}

The tank had a flat section on the top 1 m wide along the entire length where
cable penetration into the tank occurred.
The base of the tank was flat and rested on steel shielding extending
along the full length and width of the tank.
Type R1408~8 $^{\prime\prime}$ Hamamatsu phototubes (PMTs) \cite{hamamatsu}
were mounted
uniformly on the walls with a mean photocathode coverage of 25\%.
This number was calculated as the total area of flat discs of the same diameter
as the photocathode surface in the PMTs divided by the area of the
tank walls scaled to the photocathode surface.
The photocathode surface was approximately 25 ${\,{\rm cm}}$ inside the tank
wall.
Liquid entered the tank at the bottom closest to the target and
an overflow outlet was situated at the top furthest end.
Cables from the detector penetrated the tank at the top, passed outside the
top of the tank toward the upstream end, and passed under the veto shield near
the bottom front.
In discussing the experimental geometry, the coordinate system that will be
used
 is positive $z$ along the tank axis in the beam direction, $y$ vertical
and positive up.
The center of the coordinate system was at the geometrical center of the
cylinder.

\subsubsection*{3.1.2\ \ The veto shield}%

The detector was enclosed by a veto shield.
This shield covered the entire detector apart from a support structure on the
floor.
It consisted of two parts, an upstream wall which was installed first and an
approximately cylindrical part with a downstream wall which was rolled into
place on rails after the detector was fully installed.
This made the coverage of the shield as hermetic as possible.
The shield had been built for a previous experiment \cite{veto} and had liquid
scintillator as an outer layer with 12.5 ${\,{\rm cm}}$ of lead shielding
inside
the scintillator tank.
A section of the veto shield is shown in Fig. 10.

\begin{figure}[ht]
\centerline{\psfig{figure=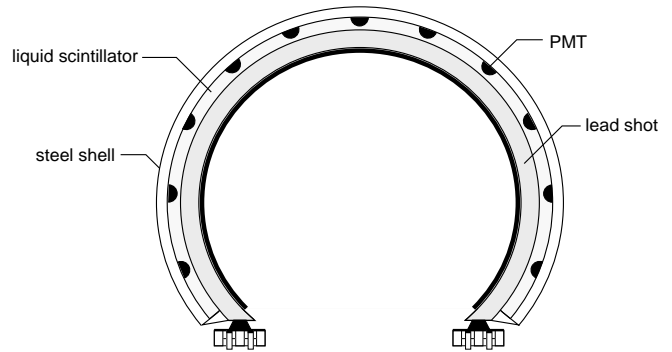,width=3.4in,silent=}}
\caption{A cross section of the veto system through the x - y plane, roughly
normal
 to the beam direction}
\end{figure}

The lead shielding absorbed neutral particles (photons and neutrons) which
traversed the active veto before reaching the detector.
Muons which stopped and decayed in the lead, producing Bremsstrahlung photons
from electrons, were tagged by the active veto.
The scintillator was viewed from the outside through portholes by a total of
292
5$^{\prime\prime}$ E.M.I. PMTs type 9870B.
Twelve plastic scintillator counters (``crack counters'') were used to cover
the optically weak
region around the joint between the upstream veto wall and the veto cylinder.
After the 1993 run, eighteen plastic scintillator counters
(``bottom edge counters'') were added along the
bottom edge of the veto counter to reject some of the cosmic rays that entered
the detector from the sides near the floor.

\subsubsection*{3.1.3\ \ Passive Shielding}%

The detector was situated 29.7 m from the production target under
2 kg/{\,{\rm cm}}$^{2}$ of steel overburden to shield from cosmic rays.
This shielding was sufficiently thick to filter the hadronic component of
cosmic rays but was penetrated by muons from cosmic-ray showers.
These penetrating muons passed through the veto shield and detector and gave a
rate in the detector of $\sim 4 {\,{\rm~kHz}}$.
Ten percent of these muons stopped in the detector and decayed, giving a
source of electrons that was invaluable for calibration purposes.
The detector was located in the veto shield inside a tunnel just large enough
in diameter to contain the veto itself.
A short portion of tunnel upstream of the detector was used for services and
partial access to the detector and preamplifier electronics.
Downstream, the detector was shielded by an 8 m long tank filled with water
that also fitted in the tunnel closely as shown in Fig. 1.
The detector rested on a concrete pad covered by six inch thick steel blocks.
The cracks between these blocks were filled with steel shot.
Upstream access was made through a 1m diameter labyrinth from the surface.
Ventilation was accomplished through a second small-diameter tunnel that also
had a labyrinthical character.
Downstream of the beam stop, shielding equivalent to 9 m of steel attenuated
the
neutron flux produced in the beam interaction process.
This resulted in a negligible neutron rate in the detector associated with the
beam.
This hermetically sealed structure provided excellent shielding both from
cosmic
rays and from beam-associated particles.

\subsection*{3.2\ \ Liquid}

The detector medium was designed to be sensitive to {$\check{\rm C}$erenkov }
light from electrons and
relativistic muons and scintillation light from all charged particles.
A dilute mineral-oil-based scintillator was used to accomplish this.
Mineral oil was well suited for this detector.
It had radioactive contamination below a part in 10$^{12}$, it retained a
stable
attenuation length as it was not exposed to oxygen, it was
non-toxic, had a high flash point (105$^{\circ}$ C),
and it had a high index of refraction of 1.47, implying a
{$\check{\rm C}$erenkov } cone angle of 47.1$^{\circ}$  for $\beta\sim 1$
particles.
This high index of refraction increased the amount of {$\check{\rm C}$erenkov }
light that was
radiated by an electron by a factor of 1.57 compared to water.
Chemically, mineral oil is composed of linear chain hydrocarbon molecules
($C_n~H_{2n+2}$)  where $n$ varies from 22 to 26.
Isotopically the carbon was 1.1\% $^{13}$C and 98.9\% $^{12}$C.

Extensive testing of various mineral-oil-based scintillators was performed in a
 beam containing positrons and protons in a test channel at LAMPF
\cite{reeder}.
Angular distributions of the light emitted by relativistic and non-relativistic
 particles in this beam allowed optimization of the ratio of isotropic light to
{$\check{\rm C}$erenkov } light.
Typical angular distributions for water, pure mineral oil and the mixture that
was used in this detector are shown in reference \cite{reeder}.
A concentration of about 0.031 g/l of b-PBD
(butyl-phenyl-bipheny-oxydiazole, CHNO) in mineral oil was determined to be
optimum for this detector.
For a fast moving particle ($\beta = 1$), the ratio of the number of
photoelectrons generated from isotropic light to that from direct
{$\check{\rm C}$erenkov } light was determined to be about 5 to 1.
 Isotropic light was emitted directly by the scintillator from particle
ionization
 together with a contribution from ultraviolet {$\check{\rm C}$erenkov } light
that was absorbed near the production
point and then reemitted by the scintillator.
A global fit to the sum of isotropic light and that in the {$\check{\rm
C}$erenkov } cone
was used for particle identification and to reconstruct vertex and angle
information for relativistic particles.

Tests were also conducted to ensure that those components that were in direct
contact with the b-PBD liquid scintillator, such as signal cables, PMT bases,
and the interior black epoxy paint, did not contaminate the scintillator.
Light attenuation measurements were periodically made at eight different
wavelengths from 340 to
550 nm to monitor the ageing of scintillator maintained at elevated
 temperatures exposed to these components.

The oil that was used was Britol 6 NF HP White Mineral Oil \cite{oil}.
The oil was delivered to the detector site in December 1992 and stored in a
52,000-gallon storage tank next to the neutrino tunnel.
This tank was protected internally in the same way as the detector tank.
After the PMTs were installed and the detector sealed, the oil
was pumped from the storage tank into the detector through stainless steel
plumbing in April 1993.
The b-PBD scintillator was added to the mineral oil inside
 the detector over a two-week period in August 1993.
A total of 6 kg of b-PBD powder was dissolved in approximately 50,000 gallons
of oil.
The final concentration of b-PBD was measured to be 0.031 g/l.
During the time of storage and while inside the detector, air was excluded and
the
liquid was exposed to a pure nitrogen atmosphere.
Samples of liquid scintillator taken from near the top and near the bottom of
the
 detector after the 1994 period showed no difference in the concentration of
b-PBD at these two extreme positions within the measured uncertainty of 5\%.

Liquid inside the detector was kept at a constant temperature of $\sim 15
^{\circ}$
 C by a heat transfer unit.
The density of the scintillator was measured to be 0.85 g/{\,{\rm cm}}$^{3}$,
giving a number density for carbon atoms of $n  = 3.63 \times 10^{22}$ {\,{\rm
cm}}$^{-3}$.
The ratios of other target atoms to carbon in the scintillator were $H/C =
2.05$,
 $N/C = 3 \times 10^{-6}$, and $O/C = 1.5 \times 10^{-6}$.
The attenuation length of the scintillator was monitored periodically during
all
running periods to determine changes that would indicate ageing or
contamination.
No deviation, within errors, from the starting attenuation lengths was observed
at the eight wavelengths monitored.
An attenuation length of about 20 m was characteristic of the scintillator
at wavelengths of $\sim$400 nm near the peak quantum efficiency for the
Hamamatsu PMTs.
The variation of attenuation with wavelength is shown in section 5.2.

\subsection*{3.3\ \ PMT system and preamplifiers}%

The PMTs that were used in the detector tank were all
tested and calibrated after delivery from Hamamatsu.
The tubes were built to a special contract using low radioactivity glass.
Testing consisted of physical inspection, base attachment, testing under
high voltage, calibration and potting.
Each PMT with base was put in a light-tight enclosure equipped with a light
pulser.
The light pulser, located 10 inches from the PMT photocathode, consisted of two
light sources, a green LED for charge tests, and an avalanche transistor for
timing
tests.
First, the PMTs sat under voltage for 8 hours with the dark rate measured every
minute.
Then charge spectra were taken at two voltages close to the suggested
operating voltage by the manufacturer.
The final operating voltage was determined by interpolation for a gain of
$\sim 4 \times 10^{6}$.
Then noise and signal rate were determined as a function of voltage to see if a
satisfactory plateau existed.
A typical pulse height distribution for single photoelectron illumination is
shown
in Fig. 11.
All of the tubes used in the detector were required to have a definite
peak-to-valley in the single photoelectron distribution as shown in the figure.

\begin{figure}[ht]
\centerline{\psfig{figure=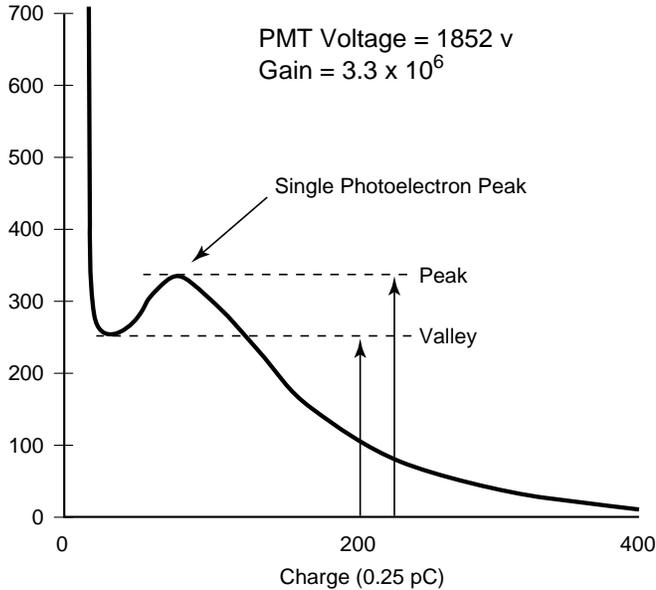,width=3.4in,silent=}}
\caption{PMT pulse height distribution for a typical phototube at normal
incidence.  The light level is such that tubes produce less than one
photoelectron on average}
\end{figure}

\begin{figure}[ht]
\centerline{\psfig{figure=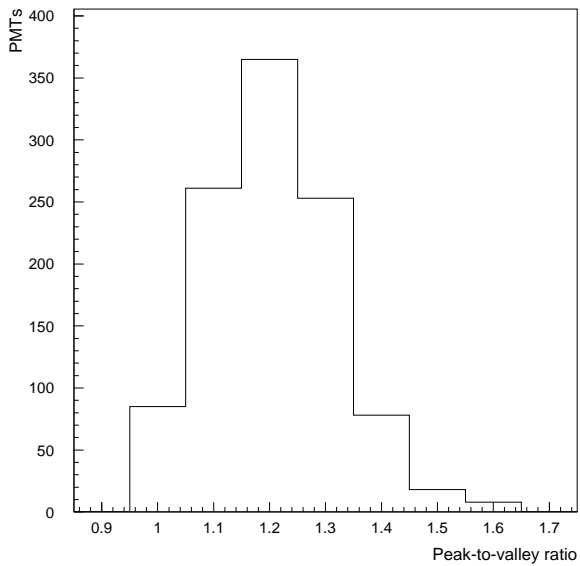,width=3.4in,silent=}}
\caption{Distribution of peak to valley parameter for detector PMTs}
\end{figure}

A distribution of the peak to valley parameter for the tubes used in the
detector
 is shown in Fig. 12.

 After testing, the base portion of the PMTs was dipped in Hysol, a
one-component
urethane material, which was primarily designed for printed circuit board
coating.
Hysol PC18 had high dielectric strength, cured at room temperature,
and was extremely abrasion and solvent resistant.
A description of the veto PMTs can be found elsewhere \cite{veto}.
The PMTs were powered through high-voltage cards that distributed to each PMT
the correct high voltage that was determined in the testing process.
These cards also provided control and monitoring of high voltage through
a monoboard computer interface.
Each PMT channel was nearly identical, consisting of a resistive base
 operated at a positive supply voltage of approximately 2000 V and stepped down
to maintain appropriate voltages at the dynodes of the PMTs used
in the detector.
A single 100-ft-long RG-58 coaxial cable connected each PMT to the preamplifier
and
 high voltage card, where the cable was terminated.
At the preamplifier input, the signal of the PMT was separated from the
supplied
high voltage on this common cable.
The preamplifier amplified the anode signal by a factor of twenty and drove
100 ft of RG-58 50$\Omega$ cable to remote electronics for charge and time
digitization.
At the input to the remote electronics after transmission through the
preamplifier
and 200' of cable, a typical single photoelectron pulse had
 an amplitude of 25 mV and a $\sim 6 ns$ rise time.
After-pulsing and undershoot on this signal were minimal.
Some additional characteristics of all the PMTs are shown in Table I.

\begin{table}
\caption{Main characteristics of PMTs}
\label{Table I}
\begin{tabular}{|cccc|}
&detector&veto&crack-counters\\ \hline
type&{\tiny HAMAMATSU R1408}&{\tiny EMI 9870B}&{\tiny HAMAMATSU R875} \\ \hline
diameter (in)&8&5&2 \\ \hline
voltage range (V)&1600-2300&710-1400&1100-1350  \\ \hline
base$~R~$(M$\Omega$)&17&15&6.27  \\ \hline
average I~($\mu$A)&110&70&190  \\
\end{tabular}
\end{table}

\subsection*{3.4\ \ Laser Calibration System}%

A  $N_2$-dye laser calibration system was installed to calibrate both timing
and
amplitude response for each individual PMT channel.
It consisted of a pulsed laser \cite{laser} located remotely from the detector.
This laser was connected to three fiber-optic cables which channeled light to
three glass bulbs $10{\,{\rm cm}}$ in diameter.
The 337 nm output wavelength was shifted by a Coumarin dye cell to 420 - 40 nm.
The time spread of the light was less than $\sim 1 ns$ rms, short compared
to the system response.
The coordinates of the flasks are given in Table II.
These glass bulbs were filled with Ludox \cite{ludox} as a dispersant.
The feedthrough bringing the light pulse into a flask entered from the top
and terminated 2 ${\,{\rm cm}}$ short of the center.
The light amplitude was varied through remotely controlled attenuators prior to
the fibers.
The average light level was varied so that all channels of the detector were
calibrated over the full response range.
Each bulb was activated independently and remotely.
The state of the pulsing system was recorded by the data acquisition system.
The positions of the bulbs were known from a survey and remained fixed
through the course of the experiment.

\begin{table}
\caption{Coordinates of Ludox bulbs}
\label{Table II}
\begin{tabular}{|cccc|}
Flask&x~(${\,{\rm cm}}$)&y~(${\,{\rm cm}}$)&z~(${\,{\rm cm}}$)  \\ \hline
1     &-36.5 &27.3 &-143.5\\ \hline
2     &35.2 &28.6 &1.3 \\ \hline
3     &-35.2 &27.9 &221.6 \\
\end{tabular}
\end{table}

\subsection*{3.5\ \ Beam Timing Signals}%

Data acquisition operated independently of the state of the beam.
This knowledge was available to off-line analysis, however, in two ways.
A prepulse was delivered from accelerator control that was used to start a
timer in data acquisition that allowed the time for each event to be recorded
with respect to this starting time.
Additionally, a level was also available that was recorded whenever beam was
present.
The mechanics of the recording process are discussed in the data acquisition
section below.

The fine time structure of the beam was determined by the RF structure present
in the 201 MHz part of the linac described in section 2.
The proton beam consisted of bunches approximately 5 ns apart (4.969) with a
width of about 200 ps rms.
Upstream of the experimental area shown in Fig. 2 were pickup plates designed
to provide a signal induced by the passage of the proton beam.
This signal consisted of pulses one nanosecond wide at the bunch frequency
201.25 MHz.
The amplitude of these signals was proportional to individual beam bunch size,
but because the beam was quite stable it proved sufficent to measure beam
arrival time with a single level discriminator.
This discriminator output was transmitted through a fiber optic link to the
remote electronics station.
At the data acquisition electronics the signal was counted down by a factor
of eight and then sixteen parallel signals were generated, each of which
repeated
after 640 ns.
Each of these signals was sent to a specially modifed digitizing channel
capable of dealing with this repetition rate at the input.
For any event, therefore, a number of measurements of beam arrival time was
available with each channel offset by $\sim$ 40 ns from its neighbour.
These measurements of beam arrival time were used to calculate the beam
arrival time relative to an event.
Knowledge of the arrival time of proton bunches was desirable for those parts
of the experiment that used pion DIF generated neutrinos.
In the study of DIF events and ${\nu_{\mu}}$C events in particular, this
information was used to identify events which maintained a time correlation
with the proton beam.
The details of the data analysis method will be described in the subsequent
publications where these data were important.

\section*{4.\ \ Data Acquisition and Calibration}%

\subsection*{4.1\ \ Overview and Architecture}%

Data were acquired by this detector on the basis of energy deposition in
a short time period.
This data acquisition system was designed to detect a primary trigger with
energy
 deposition in excess of approximately 4${\,{\rm~MeV}}$ electron equivalent and
then
to acquire data nearby in time with a different energy threshold.
In Fig. 13 is shown a schematic of a single PMT channel.

\begin{figure}[ht]
\centerline{\psfig{figure=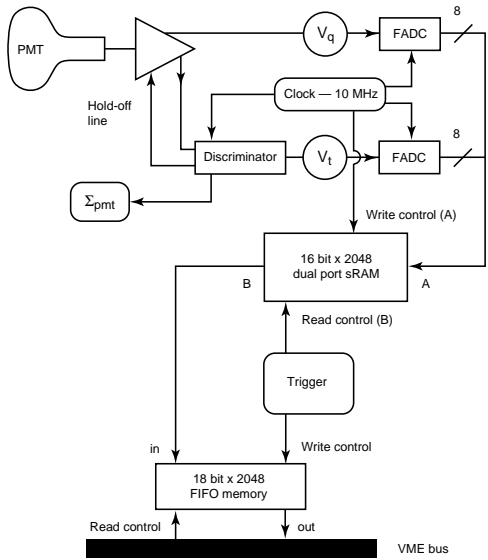,height=3.4in,silent=}}
\caption{Single channel block diagram}
\end{figure}

The electronic digitizing system was designed to collect charge from each
PMT, digitize time and amplitude, store these data for a substantial
period and generate a trigger for data retrieval.
This task was accomplished in the presence of a significant counting rate from
cosmic
rays and ambient radioactivity.
A `hit' in a channel corresponded to a minimum pulse height of $\sim 1/4$ of
the peak response to a single photoelectron in an average-gain PMT.
Neutrino induced events consisted of primary events containing more than 100
hit
tubes pulsing within 200 ns of each other, followed in time by secondary events
containing 21 to 200 hit tubes.
Secondary events occurred on timescales varying from a few hundred ${~\mu s}$
for
neutron capture to tens of milliseconds for nuclear $\beta$ decay.
The individual dark pulse rates of the PMTs ranged from 1 to 5 kHz,
producing a noise background that the acquisition system dealt with without
affecting the trigger.

The experiment ran off a local binary 10 MHz clock which was synchronized to
 a global positioning system \cite{global}.
Each channel had pulse height and fine time digitized by flash ADC \cite{fadc}
every 100 ns in synchronism with this clock.
These data were then stored in random access memory (RAM) in each channel at
 a common address, given by the least significant 11 bits of the clock.
Each channel RAM contained a $204.8{~\mu s}$ history of activity in that
channel.
A count of hit channels was updated at each 100ns clock tick for both the
detector and veto
systems and these sums were used to initiate a primary trigger and secondary
activities.
After the trigger readout order was broadcast to each channel, data were
transferred to a FIFO buffer at each channel, and data common to the event
were stored in a trigger FIFO memory.
This memory was interrogated by a master monoboard for the next level trigger
generation.
After data were transferred to channel FIFO, events were assembled at the
crate level by a local monoboard \cite{monoboard} in each crate and on
completion sent to the SGI multiprocessor system \cite{sgi} for event building
and reconstruction.
The trigger operated independently of the state of the LAMPF beam, and
 tagged individual events as in-time or out-of-time with the LAMPF proton
macropulse.
The beam-unassociated background to any neutrino signature was determined
by counting its occurrence in the beam-off sample and scaling by the
beam-on/beam-off ratio of $\sim 0.07$.
The determination of this factor is discussed in chapter six.

In the following sections some terms will be used which we define here.
A ``primary event'' was one which satisfied a threshold of $\sim 100$ hit tubes
in the
detector and had no veto signal either nearby in time or in the previous
$15.2{~\mu s}$.
An ``activity'' exceeded one of the detector thresholds without meeting
the requirements for being a ``primary''.
A ``timestamp'' was the time of a digitization as measured by the
``binary clock''
and also the address in channel memory where the ``fine time'' and ``charge''
 were recorded for each channel.
``QT cards'' housed digitizing channels for ``fine time'' and ``charge''.
Data were recorded either for a ``primary'' or an ``activity'' after a
``timestamp'' was ``broadcast'' from the trigger system to ``receivers'' in
each QT crate.

Memory pointers in each channel RAM were set to the same location, with the
address described by the least significant 11 bits of the 10MHz binary clock.
First level triggers were generated from an instantaneous global count for
either the detector or the veto shield.
This trigger produced a transfer from channel RAM to a FIFO attached
to a particular channel and a latched summary data word into trigger memory.
In parallel with this process, an event header was generated for transmission
to the multiprocessor event handling system described in section 4.4.

\subsection*{4.2\ \ Electronics}%

\subsubsection*{4.2.1\ \ Digitizing System}%

In each QT card, charge and time-of-arrival information were digitized and
stored.
The least significant eleven bits of the GPS based clock were distributed to
all modules generating times at which flash analog-to-digital converters
\cite {fadc} sampled data.
This binary number was also the address in RAM at which digitized data were
stored identically in each QT channel.
The clock was distributed from a central driver so as to arrive synchronously
at
each QT card within a few nanoseconds.
This clock determined a coarse time of arrival for PMT signals which was,
 as stated above, numerically the same as the address in the 2048 deep local
channel dual-ported RAM.
A fine arrival time for a PMT anode pulse was determined by a leading edge
comparator set to the hit threshold defined in section 4.1.
The comparator signal started a linear ramp that rose for two subsequent
``clock
ticks,'' after which the ramp was reset to zero before the next 100 ns clock
tick arrived.
Once inititated, this timing ramp was unaffected by subsequent photoelectrons
for 200 to 300 ns depending on the phase of the initial hit with respect to
the 10 MHz clock.

The ramp voltage was flash digitized at each clock tick by an 8-bit FADC
\cite{fadc} and stored in a dual-ported RAM attached to this channel.
The ``fine time'' determined this way had a least count of 0.8 ns.
The anode pulse was also connected to an integrator-stretcher that convoluted
it
with an exponential with a decay time constant of 700 ns (6000ns in 1993).
The stretched pulse was digitized by another 8-bit FADC at each clock tick, and
digital output was sent to a parallel dual-ported RAM, as part of a 32 bit
word at the identical address to the fine time.
Again, all the address counters controlling FADCs and dual-ported RAMs were
synchronized by the 100-ns clock.
At the start of each cycle through memory locations in the RAM, the addresses
of all RAMs were reset to zero to assure long term synchronization of the data
in memory.
The FADC-dual-port-RAM circuit operated continuously, filling channel
memory with the time history of that PMT.
``Latency'' will be used to describe the case in which events were retrieved
after memory was overwritten and hence contained irrelevant data.
The small loss of data from this effect is described below in section 4.3.
After a trigger decision was made (described in section 4.3), data were
transferred from appropriate addresses to a FIFO.
Interrogation took place through the second RAM port.
Data resided in this FIFO buffer until it was read out by a dedicated
 monoboard computer \cite{monoboard} servicing 16 QT cards, i.e. 128 channels.

A schematic illustration of the system layout is shown in Fig. 14.

\begin{figure}[ht]
\centerline{\psfig{figure=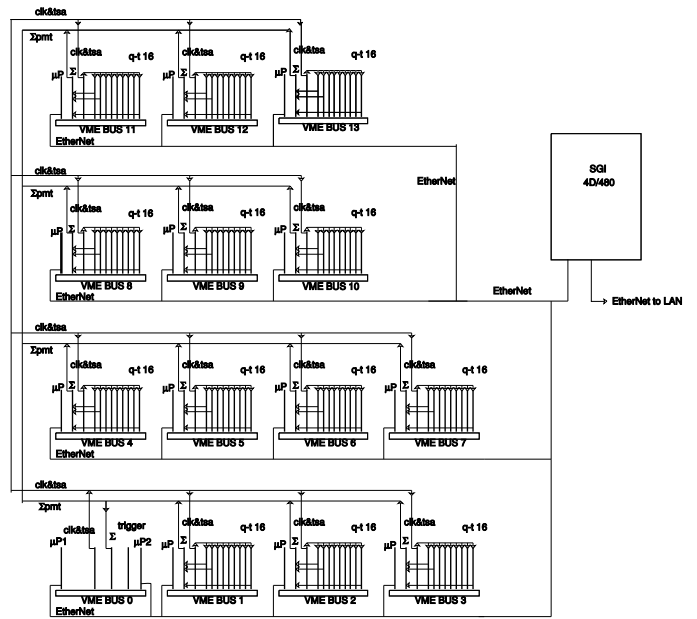,height=3.2in,silent=}}
\caption{Schematic of data acquisition system electronics}
\end{figure}

The pulse digitization system described above was built in VME format using
9-U by 380-mm cards.
This system was housed in 13 VME crates, each crate packed with
16 QT cards of eight channels per card.
In addition, each crate housed a receiver-trigger interface card, a crate
channel
 summation card, and a MVME167/D1 monoboard computer.
The receiver-trigger interface card was driven by the trigger system described
below and initiated data transfer to FIFO after receipt of a trigger broadcast.

\subsubsection*{4.2.2\ \ Channel Sum System}%

Trigger input information was derived from a majority summation system.
Each channel discriminator in a QT card generated a logic signal which was
asserted at the next clock tick after an anode pulse arrived, and remained
asserted for two clock ticks.
These logic signals were summed digitally at the crate level.
Appropriate crates were then summed digitally for both 1220 detector and 292
shield PMT signals to make an 11-bit detector sum and a 9-bit shield sum
in the central trigger crate.
These sums were available as binary signals representing a channel count of
those that fired in the last two clock intervals.
The count was updated every clock cycle.
Each binary number, detector and veto, was fed to digital comparators that
produced output if the sum was greater than or equal to values preset in
hardware.
Supplemental veto counters (``crack counters'') each generated a trigger
condition equivalent to the firing of six or more veto tubes.
These comparator signals were used as the basis of trigger decisions described
in section 4.3.
Trigger algorithms and internal operation of the trigger system are also
described in section 4.3.

\subsubsection*{4.2.3\ \ Performance Differences in 1993, and 1994/1995}%

There were several significant differences in data acquisition between data
in 1993 and 1994/95.
The transfer function in the amplitude channel of each QT card was non-linear
at low pulse heights in '93.
The gain of each channel was also somewhat different.
Full-scale for 1993 corresponded to $\approx 24$ times the peak of the
single photoelectron response (Fig. 12 for an average-gain PMT,
{\sl vs.} $\approx 30$ times for 1994/95.
In 1993, the lowest 4 ADC counts spanned $\sim 3$ times the charge that would
be expected for a linear convertor.
In '94 changes were introduced that made the transfer function linear up
to the top of the ADC range.
In '93 the integration time constant for time digitization recovery was
$6{~\mu s}$; in '94/'95 this was changed to $0.7{~\mu s}$.
In both cases the effective transfer function was equalized in higher level
software leading to very similar detector properties in each channel.
After correction, a small degradation of resolution remained due to the
large charge spread for low ADC counts.

\onecolumn
\begin{table}[p]
\caption{Table of tank hit D and veto hit V requirements for
various trigger types for 1993 through 1995 running periods.
Betas generate a single trigger broadcast.  Primaries add the broadcast
of all Activities within the previous 52.8 $\mu$sec.  In addition,
some Primaries cause the broadcast of all Gammas within the following
1 ms, and some generate the six $\mu$sec lookback.}
\label{Table III}
\begin{center}
\begin{tabular}{|c|c|c|c|c|}
1993 &1994 early  &1994 late &1995 & Function \\ \hline
$D\ge18 or V\ge 6$&$D\ge18 or V\ge 6$&$D\ge18 or V\ge 6$&$D\ge18 or V\ge 6$&
Activity \\ \hline
$D\ge 21 and V< 6$&$D\ge 21 and V< 4$&$D\ge 21 and V< 4$&$D\ge 21 and
 V< 4$&$\gamma$\\ \hline
none&none&none&$D\ge75 or V< 4$&$\beta$ \\ \hline
$D\ge100 and V< 6$&$D\ge 100 and V< 4$&$D\ge 125 and V< 4$&$D\ge150
and V< 4$&Primary \\ \hline
$D\ge 300 and V< 6$&$D\ge 100 and V< 4$&$D\ge 125 and V< 4$&$D\ge150
and V< 4$&Primary, $\gamma$ window\\ \hline
none&none&none&$D\ge 300 and V< 4$&Primary, $\gamma$, lookback\\
\hline
\end{tabular}
\end{center}
\end{table}
\twocolumn

\subsection*{4.3\ \ Trigger}%

\subsubsection*{4.3.1\ \ System Hardware}%

The information on which trigger decisions were based was an updating sum of
hit
 tubes described in section 4.2.2.
Trigger electronics and monoboards were housed in a single 6U crate.
A block diagram of this system is shown in Fig. 15.

\begin{figure}[ht]
\centerline{\psfig{figure=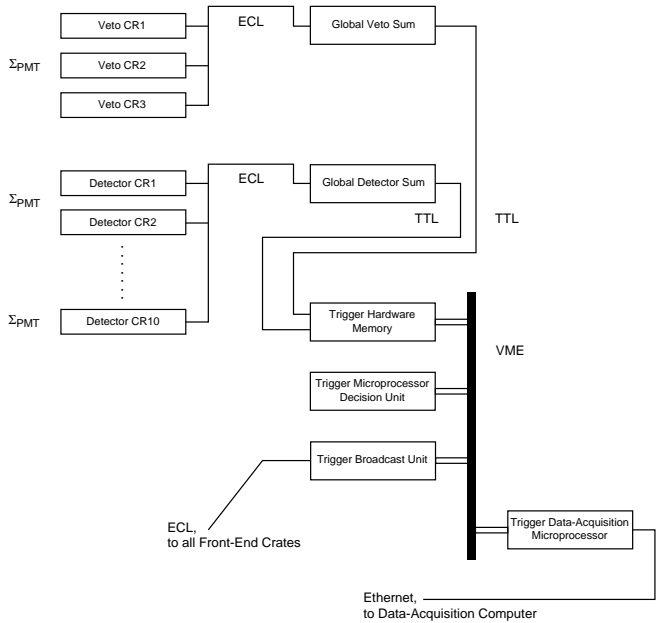,width=3.4in,silent=}}
\caption{Trigger system hardware schematic}
\end{figure}

The modules consisted of a global digital sum card, comparator and logic
memory FIFO card, two monoboards which were referred to as master and
satellite, and a broadcast card.
The detector and veto shield global digital sum cards referred to in section
4.2 completed the global sums that were used in the trigger.

These sums were interrogated by a comparator system which set levels when
global sums exceeded preset values.
These values are referred to as $D_n$ for the detector and $V_n$ for the veto.
In Table III is shown the comparator level settings for each
period of data taking.
The primary threshold was raised in the middle of the 1994 run, and this
is indicated in Table III.
After a primary or activity comparator level was set, data describing the
event were written to a trigger memory FIFO.
These data included the full 16 bit binary clock time, the state
of the comparator bits and four external data lines describing laser
status, beam gate, bottom edge veto counters and prebeam timing signal.
The trigger master monoboard polled this FIFO and, whenever data were present,
subjected FIFO data to conditions for primary event selection and for
activities
 nearby in time, both of which are described in the software section below.
On receipt of a completed event, header information was assembled by the
satellite monoboards, and event addresses were broadcast to receiver cards
located in those 13 crates that housed QT cards.
This occurred when the master monoboard loaded a register in the broadcast
card with the timestamp.
After receipt of this timestamp in the receiver cards, transfer from channel
 RAM to FIFO occurred.
Presence of a data present signal in channel FIFO prompted each QT crate
monoboard to transfer six contiguous locations in each FIFO to memory.
This process continued until all channel FIFOs were empty.
On completion, these assembled event components were transferred to the
SGI multiprocessor system through an ethernet link of all system monoboards
described in section 4.4.

\begin{figure}[ht]
\centerline{\psfig{figure=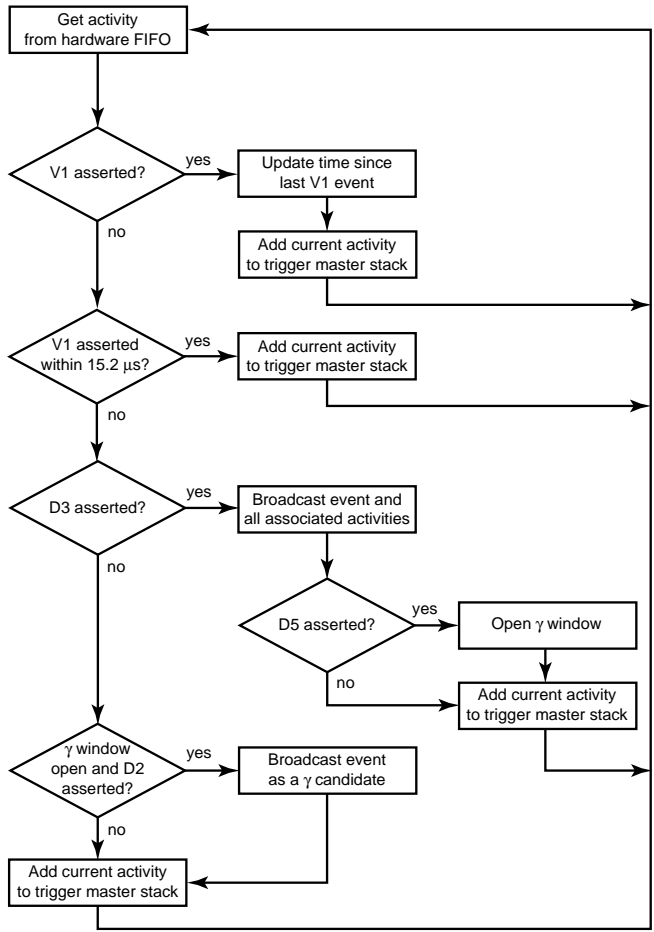,width=3.4in,silent=}}
\caption{Flow diagram, master monoboard software}
\end{figure}

\subsubsection*{4.3.2\ \ Software Selection}%

A flow diagram for event selection in the master monoboard is shown in
Fig. 16.
Input data for this process were contained in the trigger memory FIFO and
in a circular buffer, maintained in master monoboard memory, of all
activities that asserted an appropriate D or V level.
This activity data included the binary time of occurrence and
comparator levels in veto or detector.
After each of the checks described below, the activity found in trigger memory
was added to the activity stack in the master monoboard buffer.
The first loop in Fig. 16 checked for veto activity and
updated the record of the time of the previous veto event.
If there was no veto activity, a check was made for veto activity in the
previous $15.2{~\mu s}$.
If neither of these were true and the detector primary level was asserted,
the event was accepted as a primary trigger.
A broadcast was then initiated and QT data moved to channel FIFO.
If a primary trigger remained asserted after six clock ticks then a
broadcast was initiated for the next six.
This often happened after a trigger from a high energy cosmic ray event made
enough late scintillation light.
Each six ticks, with five intervals was referred to as a ``molecule''.
If in addition the detector threshold was met, a ${\gamma}$ flag was set.
If the ${\gamma}$ flag was asserted an activity in the subsequent millisecond
was classified as a ${\gamma}$ candidate.

In summary, primary triggers were determined mainly by a threshold in
the detector but with restrictions from the veto shield.
After a primary, a lower threshold was used to initiate broadcasts as
${\gamma}$
candidates.
When the event was passed on to the next level, up to four activities closest
in time to the primary in the $51.2{~\mu s}$~ previous to the event were
broadcast as well as all ${\gamma}$ candidates in the succeeding 1 ms.
The 100-tube trigger corresponded to $\sim 4 {\,{\rm~MeV}}$ in electron energy
and
the 21-tube trigger to $\sim 0.7 {\,{\rm~MeV}}$ in ${\gamma}$ energy.
A $\beta$ trigger generated a single trigger broadcast.
In 1995, a primary level with $\gamma$ was used to ``look back'' at the
${6\,\mu{\rm s}}$ prior to the primary, by generating two broadcasts to
cover that period.

\subsubsection*{4.3.3\ \ Trigger Rates}%

As was described above, the basic data rate of the system was the 10MHz clock
driving digitization rates.
The synchronous discriminator in each data channel fired at an average rate of
5${\,{\rm~kHz}}$, reflecting the sum of anode pulse rates in each channel from
dark current
 and physical processes in the detector.
  The mode of the dark current rate in the ensemble of PMTs was
2${\,{\rm~kHz}}$.
The remaining rate was dominated by cosmic ray muons and was 4${\,{\rm~kHz}}$
under the
overburden of the detector.
Either D or V was asserted at a total rate of $\sim 17 {\,{\rm~kHz}}$,
prompting address data transfer into the trigger memory FIFO.``Primary event''
 identification by the master trigger monoboard produced a rate
 of 20${\,{\rm~Hz}}$.
Each of these primaries on average had associated activities that raised the
data rate to $\sim 40{\,{\rm~Hz}}$, which was the rate of broadcasts to QT
crates for
data accumulation at crate level.
This trigger rate of $\sim 40{\,{\rm~Hz}}$ was dominated by a few
beam-unrelated
processes.
Beam-related neutrino interactions occurred at a rate of only about one per
 hour, which is down by a factor of $\sim 10^{-5}$ compared to the beam-off
background.
Fig. 17
shows the raw PMT multiplicity distribution for all broadcast events.

\begin{figure}[ht]
\centerline{\psfig{figure=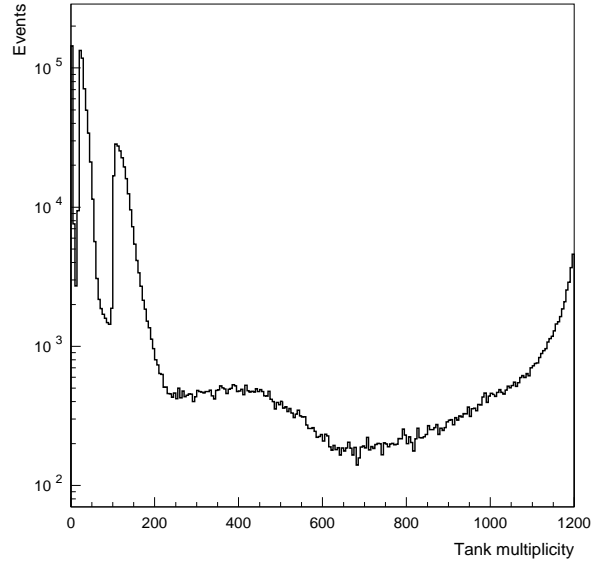,width=3.4in,silent=}}
\caption{Raw PMT multiplicity distributions for all broadcast events}
\end{figure}

The high end of the spectrum, where almost every PMT was hit,
corresponded to through-going cosmic muons.
The region below the peak, from 650-1050 hit PMTs, was dominated by
a combination of cosmic-ray muons and neutrons.
Michel electrons from cosmic muons that stopped and decayed in the detector
occurred in the 250--650 hit PMT region and were greatly
reduced by disabling the trigger for $15.2{~\mu s}$ following activity in
the veto shield.
Finally, electrons from $^{12}$B $\beta$ decay, produced dominantly by
cosmic-ray
$\mu^-$ that captured on $^{12}$C, typically had 100-250 hit PMTs while
 radioactive~${\gamma}$ (e.g., from Th, K, and Co decay) have $<100$ hit PMTs.
There is a sharp edge at 100 hit PMTs in Fig. 17 because the $D_3$
level is set to 100 for these data.
 Of the total rate, approximately 10${\,{\rm~Hz}}$ were beta decays of $^{12}$B
from muon capture,
 10${\,{\rm~Hz}}$ from natural radioactivity in the PMT glass and surrounding
material,
 0.11${\,{\rm~Hz}}$ from decay electrons from stopped cosmic ray muons,
1${\,{\rm~Hz}}$ from muons
 that
evaded the veto shield and crack counters and 1${\,{\rm~Hz}}$ from neutrons
produced
from muons that did not enter the detector.
These were the rates of primary triggers; associated activities contributed a
similar rate.

\subsubsection*{4.3.4\ \ Latency, FIFO Half Full Inhibit and Trigger
Efficiency}%

Data taking rates in the detector were quite constant because they were
completely dominated by cosmic rays and natural radioactivity.
The trigger rate was adjusted to an acceptable level at the start of
data taking by varying the settings of the D$_n$ comparators for primary
threshold and for activities.
Occasionally, fluctuations in these rates caused the channel FIFOs to fill up,
and when the half full flag was asserted in any FIFO the trigger was inhibited
until the situation was cleared by data acquisition.
This occurred $< 1\%$ of the time.
Similarly, the master trigger monoboard occasionally took long enough to
process
an event that channel RAM was overwritten, a condition referred to as latency.
The loss of data from latency was $\approx 1\%$.
These losses remained approximately constant throughout data taking as
might be expected from the primary sources of data in the detector.
One source of these conditions occurred when the SGI multiprocessor did
not remove data from crate monoboards, whereupon channel FIFOs were not
emptied and the acquisition part of the system was halted.
This effect was included in the FIFO half full estimate and is the major source
of this condition.

\subsection*{4.4\ \ Data Readout and Event Building}%

\subsubsection*{4.4.1\ \ Raw Data}%

The primary trigger caused data to be transferred from channel dual-port
RAM to channel FIFO in sets of six timestamps.
This corresponded to a time window of $0.5{~\mu s}$.
Each crate monoboard running asynchronously was set up to poll all FIFOs
in the crate to determine if data were present and if so, raw data were
transferred to monoboard memory.
Two bytes per timestamp (one each for time and charge digitizers) were
read out for each channel in a crate.
For a full crate of 128 channels, 1536 bytes of raw data were read out
over VME for a six timestamp event.
In each crate monoboard, a 32-bit header was constructed by interrogating
the crate receiver card.
This header included 11 bits of timestamp address (TSA) and an 8 bit
counter to assure FIFO synchronization.

\subsubsection*{4.4.2\ \ Compact Data}%

Compact data were formed in the QT monoboards from raw data.
Information was reported from six sequential TSAs covering five time intervals
for hit tubes in the crate.
Longer events were handled in a slightly more complex way, but the essential
point
was that six timestamp molecules formed the basis for data transmission.
Each compact data block was made up of a header followed by channel data
for tubes that were hit.
The compact data header included information which allowed the event builder to
verify proper assembly of event fragments.
Channel data consisted of a 16-bit header, followed by 8 bits for each
interval with fine time and 8 bits for each charge deposition.
After an interval with fine time information, the QT hardware did not
provide fine time information for the next two intervals, but charge
increments were still registered.
Baseline amplitude was the amplitude before pulse height digitization and
was used for baseline subtraction.
An interval stored with fine time was termed a ``hit with fine time''.
If no fine time was available it was called a ``hit without fine time''.

In each crate monoboard, data were searched for hit PMTs by looking for
 rise and fall of the timing ramp as a function of time as signature.
The difference between the baseline and maximum value of timing ramp was
stored for later determination of fine time.
For a hit occurring between TSA n and n + 1, fine time information was
stored as ADCt(n + 2) - ADCt(n).
In the last interval, however, fine time was calculated using a default
 value for the timing slope.
To first order, the PMT charge $V_q$ in a given interval was simply the
difference between ADC measurements before and after that interval.
This was complicated by 3 effects:
(1) $V_q$ required at least 30 ns to attain maximum value;
(2) $V_q$ was slew-limited at 2.2 ADC channels per ns (this was only relevant
for hits
over 66 ADC counts);
(3) $V_q$ decayed to its original baseline voltage approximately as an
exponential
with $\tau$ = 755 ns (6000ns in 1993).
These effects were taken into account in computing the
charge for each hit, using a linear approximation to the exponential
effect (3).
Except for 1993 data, additional charge increments occuring close
enough in time to a hit with fine time were ascribed to effects (1) and (2),
 and included with that hit.
If the charge ADC was set to full scale immediately before a hit, the
overshoot bit was set in the ``baseline'' variable for that PMT channel.
The remaining 5 bits of the baseline were set to the 5 most significant
bits of the ADC.
These bits allowed a determination of ADC saturation.

\subsubsection*{4.4.3\ \ Event Building and Calibrated Data}%

PMT data were transferred to a multiprocessor system \cite{sgi}
over two independent ethernet segments.
Both raw and compact data were transferred to check integrity of compact
data, but during the operation of the experiment raw data were prescaled
to avoid acquisition induced dead time.
In the multiprocessor, individual timestamp molecules from each crate
were ordered by the CODA \cite{coda} package.
Complete events were then built using data from the first four time
stamps in each molecule.
Apart from 1993 data, when these events were being built, the time
information was converted to times relative to the first timestamp of
the event; and each PMT time and charge was calibrated using constants
determined from laser events, as described in section 4.6.
The SGI multiprocessor contained 8 independent processors.
Events were built in one processor, a second distributed events to the
remaining six processors that were used exclusively for event
reconstruction on line.
The output of event building was the Calibrated Data bank in ZEBRA format,
 which served as primary input to the analysis package used both online and
offline \cite{ion_chep} .
A header included a calibration data file identifier, number
of timestamps in the event, and a hit count with and without fine
time.
This header was followed by tube number, charge and time and saturation
for each hit.

\subsection*{4.5\ \ Corrections for Digitizer Charge Response, 1993}%

For '93 data, the Q circuitry had a $6{~\mu s}$ decay constant, and effects
(1) and (2) above were not taken into account by the online software.
The result was that in the '93 data most hits were accompanied by hits
without fine time in the subsequent interval.
This effect was dealt with offline by adding the charge from these
subsequent hits into the charge for the first hit.
Another effect unique to the '93 data was a nonlinear relationship
between PMT charge and ADC value (especially noticable at low charge
levels).
This problem was also dealt with offline using a nonlinear mapping to
recover PMT charge.
After the '93 run the nonlinearity was removed by a hardware upgrade.
The quality of the 1993 data was high after correction.
For example, the energy resolution deduced from Michel electrons was
as good in 1993 as in 1994 and 1995.

\subsection*{4.6\ \ Laser Calibration}%

Individual channel calibration was performed using the pulsed laser
system described in section 3.4.
A laser induced event was subjected to the same requirements ($> 100$ hit
tubes) as for any primary trigger.
However, each event was identified by a set bit in fixed data in trigger
memory.
The laser system was pulsed at $\sim 0.1 {\,{\rm~Hz}}$ continuously and
asynchronously with the accelerator during normal data taking and at
20--30 ${\,{\rm~Hz}}$ during beam off periods for special purpose runs.
This operation induced a negligible dead time, while providing a
stroboscopic probing of the detector at times that were random with
respect to beam arrival.
The laser pulse time was tagged as a special event by the data
acquisition system.
Gain calibration was obtained from low light intensity runs by fitting
the resolved single photoelectron peak for each PMT individually.
Calibrations were performed periodically during the runs and have shown
that calibration constants remained stable throughout the data-taking
period.
This low level of illumination was not adequate to provide timing
calibration, so special data runs were interspersed at high light
levels to provide timing offsets for all channels.
Time-offsets of each channel were calculated individually from these
high-intensity runs alone.
Time-slewing effects were assumed to be identical for all PMTs, and a
 mean effect was calculated using laser data from all tubes, resulting
in times calibrated to $\sim 1 $ ns for each PMT channel.

\section*{5.\ \ Detector Simulation}%

A detailed Monte Carlo simulation, LSNDMC\cite{lsndmc}, was written to simulate
events in the detector using GEANT.
The goal of the simulation was a complete description of events in the
detector.
For the decay-at-rest analysis, data from Michel decays of stopped muons and
from captured cosmic ray neutrons were used to calibrate directly the response
of the tank to the positrons and photons of interest.
The simulation was used in this analysis to test understanding of the detector,
particularly physical processes in the tank, reconstruction algorithms, and
 the computation of backgrounds.
For the DIF ${\nu_{\mu}}$ to ${\nu_e}$ oscillation analysis, with higher energy
electrons,  the simulation was additionally used to find the efficiencies of
analysis cuts.
This implies greater sensiivity of the results to simulation details.
This chapter focuses on issues of importance for the DAR analysis, and on
using the simulation to understand physics processes occurring inside
the tank.
Some comparisons to Michel data will be presented in section 6.3.

\subsection*{5.1\ \ Monte Carlo Ingredients}

The GEANT geometry was defined to include the main detector tank, veto shield,
and overburden for detector shielding.
The latter two portions were implemented in order to track cosmic ray muons.
Tank PMTs were positioned with a regular spacing, and within
a few ${\,{\rm cm}}$ of the actual detector locations.
They were represented as having hemispherical photocathodes, although when a
computation of optical reflection probabilities was needed, an ellipsoidal
shape could be substituted.

The Monte Carlo package included a variety of generator options \cite{lsndmc}.
Cosmic ray muons were generated with energy and angle distributions
appropriate for the altitude of LSND.
Among the variety of generator options included in the Monte carlo package
\cite{lsndmc} was a cosmic ray facility.
Each muon was tracked through the overburden and veto shield, and so
a set of starting locations for the generation of decay (Michel) $e^\pm$ was
obtained.
This allowed data {\sl vs.} Monte Carlo comparisons.
Also, estimates of the total muon rate in the tank ($\sim 4{\,{\rm~kHz}}$) and
the stopped muon rate ($\sim 0.35{\,{\rm~kHz}}$) were in rough agreement
with observations.

This simulation is addressed in detail in the following section.
 {$\check{\rm C}$erenkov } and scintillation light were simulated in detail,
and each optical
photon was tracked until it was either absorbed or generated a photoelectron.
The PMT response to this photoelectron was also simulated.
This set of steps resulted in one ``hit'', in the GEANT sense, per
photoelectron.

The digitization electronics and the online data acquisition algorithms were
 simulated using the full set of hits, yielding a time and an amplitude
for each hit PMT.
The hit time was simulated by using PMT pulse shapes determined from bench
tests and laser data; pulses for multiple photoelectrons in the same tube
were summed.
The amplitude simulation used so far was based upon the 1993 version of both
the QT digitizer cards and data acquisition algorithms,
 including the non-linear digitizer charge response.
  This non linearity was taken out in calibration.
Both the simulation and the calibration of this response were based on bench
measurements.
Because the quantization imposed by digitization occurred for
the non-linear amplitude, the 1993 data {\sl vs.} Monte Carlo comparison was
a more stringent test than the same comparison for 1994/95 data using QT cards
 that were linear.
The digitizer time response depended upon details of pulse shapes, thresholds
and fluctuations, and had a noticeable impact upon the way reconstruction
algorithms worked.
The discriminator in each channel produced a dead time after firing that
inhibited measurements of subsequent photoelectrons.
This effect was simulated in detail since it had a significant effect on actual
 event timing distributions.

\subsection*{5.2\ \ Simulation of Photoelectron Production in the Tank}%

The basic physical mechanisms which occurred in the LSND medium and which
contributed to a detected signal were scintillation from ionization
energy deposits by charged particles and {$\check{\rm C}$erenkov } radiation by
charged particles
with $\beta > 1/1.47$, including a contribution from $\delta$ rays.
Optical photons from these processes were transmitted through the medium,
taking into account absorption with or without fluorescent
reemission.
Photons could be reflected from the upper surface of the liquid or at the PMT
surface.
Photoelectron emission occurred at a photocathode and the spread from the
multiplication process internal to a PMT was also simulated.
Charged particles in the medium were tracked using GEANT, including particle
interactions and full electromagnetic showers down to cutoff kinetic energies
 of $10{\,{\rm~keV}}$ for~$e^\pm$'s and ${\gamma}$s.
In order to simulate the contribution of {$\check{\rm C}$erenkov } radiation
from $\delta$ rays, a minimum $\delta$ ray kinetic energy of $180{\,{\rm~keV}}$
 was used.
Delta rays contributed several percent of the total light from an electron
event.

As GEANT stepped particles through the detector, it provided an energy loss
$\Delta E$ to ionization for each step $\Delta x$.
In the LSND simulation the mean number of scintillation photons was computed
from Birk's law \cite{birks}:
     $${\bar N}_s = {A\,\Delta E \over 1 + k_B\,\Delta E/\Delta x}\ \ \ ,$$
where $k_B$ is a saturation parameter, discussed below.
For particles with $\beta$ above the {$\check{\rm C}$erenkov } threshold
(1/1.47), the mean number
of {$\check{\rm C}$erenkov } photons in a given range of wavelength was
computed by the
formula in reference \cite{pdg}.

There are unfortunately no published measurements of $k_B$ for mineral oil
scintillators.
In the simulation we have used a value of 0.017 ${\,{\rm cm}}/MeV$, based upon
typical
measurements (scaled by density) for anthracene and several plastic
scintillators \cite{kb}.
However, other data give varied results, and there is some evidence based
on limited measurements done for mineral oil by the LSND collaboration that
the value could be as much as a factor of two smaller.
We regard $k_B$ as uncertain at this level.
This uncertainty has no effect on the response of the scintillator to
electrons of interest, but is quite relevant for protons, especially for
recoil protons in charged current neutrino interactions, for which the
$\Delta E/\Delta x$ term in the denominator can dominate.
For the lowest energy muons detectable as activities from $\nu_\mu + C$
interactions, the resulting change in light yield (and hence in
energy scale) can be as much as 20\%.

The absorption and emission properties of the dilute solution of b-PBD in
mineral oil were key ingredients in modelling the behavior of the detector.
b-PBD emitted light almost entirely at visible wavelengths ($\lambda >
350{\,{\rm~nm}}$),
 with a sharp cutoff at shorter wavelengths\cite{reeder}.
The attenuation length for a sample of the mineral oil used in the tank
was measured to be in excess of 15$\,$m for visible wavelengths, falling to
$<1\,$m for $\lambda < 335{\,{\rm~nm}}$ as shown in \cite{reeder} and discussed
 in section 3.2.
Most of the scintillation light reached a photocathode or edge of the
tank without absorption.
The same was true for {$\check{\rm C}$erenkov } light emitted at visible
wavelengths, which thus retained directionality.
{$\check{\rm C}$erenkov } light emitted in the UV, however, was absorbed in a
short distance.
Light absorbed in the mineral oil was lost, but that absorbed by the b-PBD
was reemitted isotropically via fluorescence with a probability of
0.83 and with an emission spectrum that was the same as for scintillation
light.
Hence for particles above {$\check{\rm C}$erenkov } threshold, a substantial
part of the isotropic
light was actually ``converted'' {$\check{\rm C}$erenkov } light.
This Monte Carlo simulation included mineral oil and b-PBD absorption
separately, so that the interplay of their absorption probabilities was
 properly taken into account at all wavelengths.
The combined absorbtion length for light that is not reemitted as a function of
wavelength for the LSND scintillator mix is shown in Fig. 18.
{$\check{\rm C}$erenkov } photons were not tracked below $\lambda <
350{\,{\rm~nm}}$
but were either absorbed or converted immediately.

\begin{figure}[ht]
\centerline{\psfig{figure=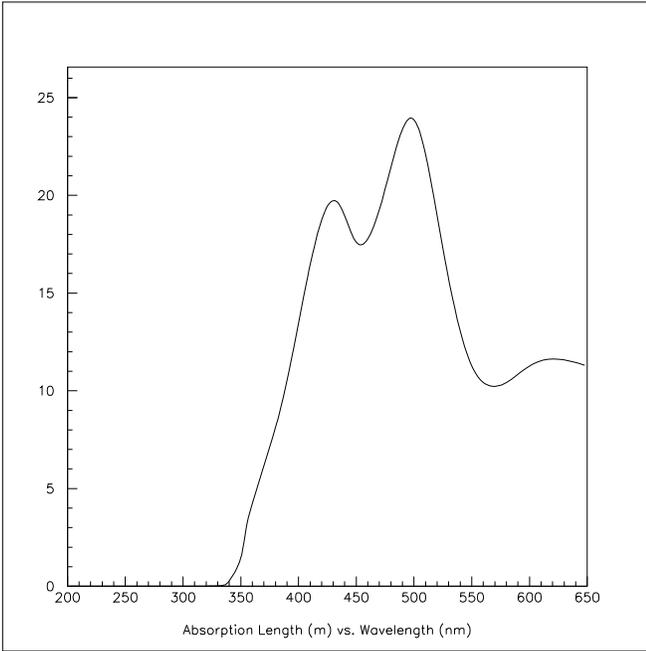,width=3.4in,silent=}}
\caption{Attenuation length for light that is absorbed without reemission
as a function of wavelength for the LSND scintillator mix}
\end{figure}

There were three components to the produced light in the tank:  direct
scintillation light, directional {$\check{\rm C}$erenkov } light, and
converted {$\check{\rm C}$erenkov } light.
The latter two components were computed with no free parameters.
To obtain the normalization constant $A$ for direct scintillation light,
measurements were performed in a test beam using the same type of
scintillator as was used in the LSND detector but with different phototubes
\cite{reeder}.
In the test beam a measurement was made of the ratio of light in the
 {$\check{\rm C}$erenkov } cone to isotropic light for $\beta\approx 1$
positrons.
This ratio was corrected back to the production point by correcting for
relative PMT response and attenuation for the two types of light.
Additional adjustments were made to conform to the precise definitions of
the three components and for a slightly different b-PBD concentration in
 the detector from the test beam.
The result was an estimate of 360 direct scintillation photons produced
per ${\,{\rm cm}}$ by a fast
$e^\pm$, along with calculated values of 325 photons per ${\,{\rm cm}}$ in the
{$\check{\rm C}$erenkov }
cone ($\lambda > 350{\,{\rm~nm}}$) and 237 photons per ${\,{\rm cm}}$ converted
from lower-wavelength {$\check{\rm C}$erenkov } photons (with $\lambda < 350
{\,{\rm~nm}}$).
The value of $A$ was set to match 360 photons per ${\,{\rm cm}}$, but
with a small correction factor determined from comparisons to real tank
data.

The actual numbers of isotropic and directional photons were selected from
a Poisson distribution with distribution means described above; each
individual photon
was tracked until absorbtion or production of a photoelectron occurred.
Absorption with reemission was also possible en route, particularly for the
shorter wavelengths; this occurred for $\sim 5\%$ of the initially
directional light.
When reflections occurred, the simulation averaged over polarization.
The {$\check{\rm C}$erenkov } cone light was spread by each of these effects.
Reflections at the PMT were modelled together with quantum efficiency
by applying the model to a wavelength-dependent complex index of
refraction \cite{moortan}
for a bialkali photocathode layer deposited inside the glass window
and including the external mineral oil and internal vacuum.
The photocathode thickness was found to be $28{\,{\rm~nm}}$  using bench tests
at $\lambda = 632{\,{\rm~nm}}$.
Phototube and top-surface reflections added $\sim 5\%$ and $\sim 4$ - 7\%,
respectively, to the total detected light.
Finally, quantum efficiency was tied to PMT measurements in air
\cite{hamamatsu}.

Prior to digitization, the amplitude response of a PMT to an emitted
photoelectron was simulated, using a model similar to Fig. 11 but with
a constant level replacing the low end peak.
This was checked for the tank itself from ultra-low intensity laser data, such
that PMT hits nearly always involved just a single photoelectron.
The match was quite close for '94/'95 data; and after inclusion of the
nonlinear
QT response, the match was also quite good for '93 data.
These comparisons also allowed calibration of the mean gains for use in
simulations.

\begin{figure}[ht]
\centerline{\psfig{figure=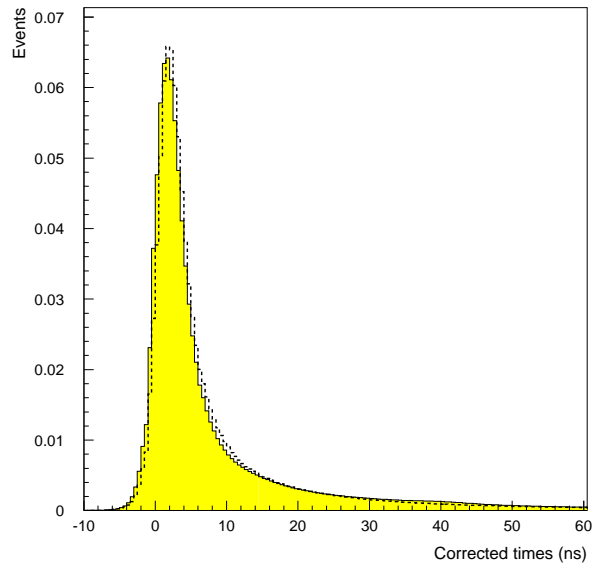,width=3.4in,silent=}}
\caption{Corrected time distribution for Michel electrons from data (solid)
and simulation (dashed) For $dist > 35 {\,{\rm cm}}$.}
\end{figure}

In Fig. 19 is a comparison of the hit time distribution for Michel events
 and the results of the simulation corrected for the event time and
time-of-flight.
These comparisons allowed fine-tuning of the parameter $A$, which in turn
allowed
 estimation of the number of photoelectrons detected per unit $e^\pm$ energy.

A main ingredient of the photoelectron simulation was a model for the time
distribution of emitted light.
Cerenkov light was emitted promptly, but scintillation light was represented
as having both fast and slow components \cite{reeder}.
Fluorescent reemmission was assigned the fast component only.
A variety of resolution effects were taken into account either explicitly or
implicitly via digitization and time calibration.
The scintillation time parameters were then adjusted to match Michel electron
data (see section 6.3.1) for PMT's with single photoelectrons away from the
{$\check{\rm C}$erenkov } cone.
Apart from a small (arbitary) offset, agreement is good even though this
does not necessarily mean that the underlying model is fully correct.
Further comparisons to Michel data are addressed in sections 6.3 and 6.4.

\section* {6. \ \ Event Reconstruction}%

\subsection*{6.1\ \ Overview}%

The event reconstruction algorithm is discussed in section 6.2.
In section 6.3 the extraction of the sample of electrons from muon decay in
the tank is discussed together with the performance of the reconstruction
algorithm on this sample.
Section 6.4 is devoted to detector energy calibration using Michel electrons.
Particle ID for neutrons and electrons is discussed in section 6.5.
In section 6.6 the determination of the duty factor and its application to beam
off subtraction is discussed.

\subsection*{6.2\ \ Event Reconstruction Algorithms}%

The event reconstruction proceeded in four stages.
The times of struck PMTs were adjusted by an algorithm which found the
most appropriate time of a nearby cluster of tubes.
Then an event vertex was reconstructed, followed by a search for a {$\check{\rm
C}$erenkov }
cone and finally verification that the distribution of times corrected to the
vertex of the hit PMTs was appropriate.

The first step in the reconstruction process was to adjust the times of all
PMTs with a pulse height that corresponded to fewer than 4 photoelectrons.
The time of each of these PMTs was compared to the four nearest neighbors and
 the time of each channel was set to the earliest of these.
This took account of the fact that signals with a small number of
photoelectrons
were distributed in time through the time spread of scintillation light
and PMT dispersion so that the normal time slewing correction was ineffective.
The use of this procedure improved the fit to the event appreciably.
An initial estimate for the event time was made as the mean PMT time
less 11 ns.
The geometrical position of each struck PMT was then corrected to a
position normal to its surface by a distance $v(t_i-t_o)$, where $t_i$ was the
PMT adjusted time, $t_0$ was the event time initial estimate, and
$v=20{\,{\rm cm}}/ns$, the velocity of light in the liquid.
The mean corrected position of the struck PMTs weighted by the square of
the pulse height was used as an initial estimate of the event position.

A $\chi_r$ function was then formed as
$$\chi_r = \sum_i (t_i-t_o-r_i/v)^2\times
W_i / (Q-6),$$
 where i ran over all struck PMTs, $r_i$ was the distance from the
PMT to the event vertex, $W_i$ was set to unity if $(t_i-t_o-r_i/v)$ was
negative and to 0.04 if it were positive.
$Q$ was the number of photoelectrons in the event.
The quantity $$(t_i-t_o-r_i/v)^2\times W_i$$  was limited to 1.5~ns$^2$.
This weighting emphasized prompt {$\check{\rm C}$erenkov } light over
scintillation light which
tended to be delayed.
An iteration was performed over $\pm x$, $\pm y$, $\pm z$ and $\pm t$ for
the vertex first in steps of 25 ${\,{\rm cm}}$ (with an equivalent time using
the velocity
v above), and then 10 ${\,{\rm cm}}$ and finally 5 ${\,{\rm cm}}$ steps.
If $\chi_r$ decreased at any iteration, then the vertex position was changed
to the new value for the next iteration.
The value of this $\chi$ was expected to offer discrimination between single
short tracks and the more complicated pattern expected from neutrons.

An angle fit was performed for each event by constructing a
$$\chi_a = \frac{1+d/4}{ 2(Q-2)}
\sum_i {(\theta_i-47.14^o)^2\over 2\times (12^o)^2}\times W_i\times
\exp(r_i/14.9),$$ where $d$ was the distance of the event vertex from the
center of the tank in meters, $Q$ was the number of photoelectrons in the
event,
and $\theta_i$ was the angle between the iterated direction
and the vector from the fitted vertex and the tube i.
$W_i$ was the PMT pulse height and $r_i$ was the distance from the
 vertex to the $i^{th}$ tube.
An initial event time for this angle fit was set at 5 ns less than the starting
 time for position.
A net of 26 angles was set up uniformly in $4\pi$.
The minimum $\chi_a$ in this net determined an initial angle,
a scan was then made in steps of 0.75 radians along the $\pm \theta$ and
$\pm \phi$ directions to find a local minimum.
The value of $\chi_a$ at this minimum was retained as a particle ID parameter.
The ratio ${(\theta_i-47.14^o)^2\over 2\times (12^o)^2}$ was limited to 0.894
 except in the ``near'' {$\check{\rm C}$erenkov } region $\cos \theta_i >
0.927$ and
$0.309 > \cos \theta_i > 0.052$, where it was set to 1.044.

The final discriminant in particle identification was the fraction of light
emitted late in the event.
Particles emitting {$\check{\rm C}$erenkov } radiation produced significant
prompt light.
A $\chi$ was constructed as the fraction of light after 12ns and was referred
to as $\chi_t$.

Each of these $\chi$ parameters was sensitive to particle type through a
measure of the complexity of the vertex, the existence of an identifiable
{$\check{\rm C}$erenkov } ring and the time distribution of the light depending
 on {$\check{\rm C}$erenkov } light and the specific ionization of the
 particle in the event.
A product was formed of the three $\chi$ as $$\chi_{tot} = 30 \times \chi_r
 \times \chi_a  \times \chi_t.$$
This parameter was effective in separating particle types.
Quantitative tests of performance are discussed in section 6.3 on Michel
electrons from muon decay and in section 6.5 on neutrons from cosmic rays.

\subsection*{6.3\ \ Michel Electron Sample}%

\subsubsection*{6.3.1\ \ Selection and Characteristics}%

Events that had veto activity in the $51.2{~\mu s}$ period previous to a
primary event were dominated by decay electrons (Michel) from both positive and
negative cosmic-ray muons that stopped in the detector.
The time distribution of these events with respect to the veto signal is shown
in Fig. 20.

\begin{figure}[ht]
\centerline{\psfig{figure=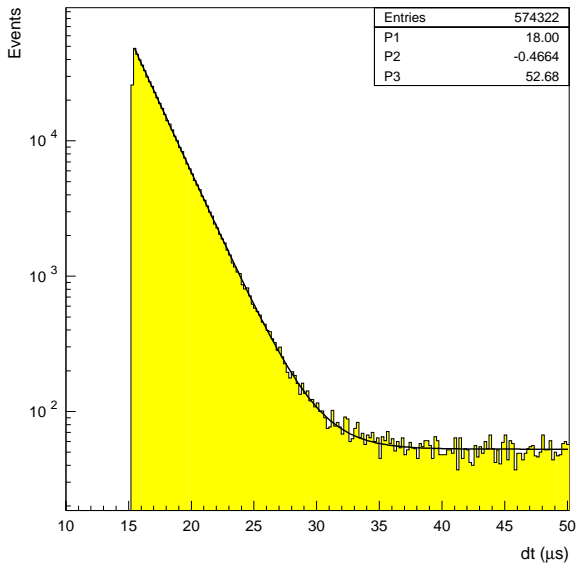,width=3.4in,silent=}}
\caption{Distribution of times of tank events after veto activity.  The
constant
P2 in the figure is the inverse of the lifetime of muons that remain after
$15.2{~\mu s}$}
\end{figure}

Few events appear in the first $15.2{~\mu s}$ after the veto event because of
an
initial trigger condition which ignored events with veto activity in this time
period, but there are still many events for analysis after this.
Events were selected to be after $15.2{~\mu s}$ from the primary trigger in
Fig. 20.
The time distribution after $15.2{~\mu s}$ was fit with a functional form of a
constant plus exponential with both amplitudes and the exponent variable.
A satisfactory fit was obtained with a value of $2.15{~\mu s}$ for the
exponent.
This agreed with the expectation for a mixture of $\mu^{+}$ and $\mu^{-}$
(which has a $2.03{~\mu s}$ lifetime in mineral oil) for cosmic rays under the
 overburden existing in this experiment.
The event sample from $15.2{~\mu s}$ to $30{~\mu s}$ after the primary has less
than
 1\% accidentals as determined by the fit to the entire $50{~\mu s}$ range.
This sample of $0.5\times 10^6$ events was used to map the gain of the detector
as a function of position and angle throughout the volume, as is discussed in
section 6.4 below.
The response of the detector to particle ID was also mapped.
The particle ID process is described in section 6.5.
 It should be noted that some accidental events in this sample were
electrons from muons that decay but that had the decay muon misidentified.

\subsubsection*{6.3.2\ \ Position Accuracy}%

 The reconstruction process provided an event position via the $\chi_r$
minimization described in section 6.2.
While there was no available direct measure of the accuracy of the position
reconstruction, the distribution in the tank of reconstructed Michel
electrons provided indirect evidence of reconstruction accuracy.
Although it was not possible to determine the distribution of stopping muons
 in the detector to sufficient accuracy for this check, it was possible to
verify that the distribution through the tank was smoothly varying and
almost uniform.
Additional evidence was available from $\nu_e C \to e^- X$ events, with a
true distribution given by the $\approx 1/r^2$ falloff of DAR neutrino
flux from A6.
For both samples, Monte Carlo simulation of the electron provided the
offset of the mean light-emission location from the point of
origin of the electron.

A single variable, $dist$ was defined as the closest distance of an event
to a surface passing through the photocathode surface of PMTs.
Events behind this surface were assigned negative $dist$ values.
In the LSND physics analyses, $dist$ was used to impose fiducial cuts.
The cut values were chosen both to help suppress cosmic ray backgrounds,
and to avoid the outermost region in which the tank response varied most
 rapidly (see section 6.4).
Distributions of $dist$ on the known samples described above have shown that
reconstructed electron positions tended to be systematically shifted away
from the center of the tank (to lower $dist$) with respect to true positions.
Quantifying this effect in the form of an acceptance correction for a given
 $dist$ cut required understanding the low-$dist$ regions well enough to
estimate the consequences of selection cuts on these known samples.
For a minimum $dist$ cut of 35 ${\,{\rm cm}}$, the resulting  acceptance
correction
(ratio of the number of events reconstructed within the cut to the true
number within the cut) ranged from 0.76 to 0.86.
If this is due to a systematic shift, then for $dist$ near 35 ${\,{\rm cm}}$
the apparent shift of an event was about 15 ${\,{\rm cm}}$.

An estimate of reconstruction {\sl precision\/} was obtained with neutrino
 events from $\nu_\mu C \to \mu^+ X$.
Both electrons and stopping muons were reconstructed for position, and the
distance between the muon and the beginning of the electron track were
compared.
A distribution of this distance is shown in Fig. 21.
This distribution agrees with the Monte Carlo simulation and gives
 an approximate vertex resolution of 20 ${\,{\rm cm}}$ for each particle.

\begin{figure}[ht]
\centerline{\psfig{figure=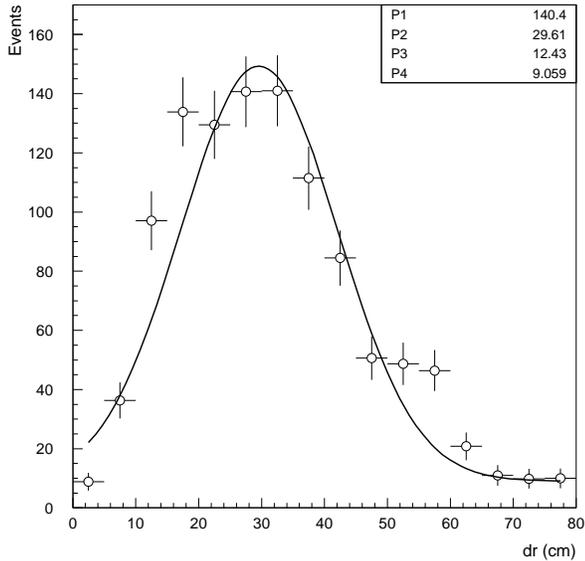,width=3.4in,silent=}}
\caption{Reconstructed distance between muon and decay electron for
${\nu_{\mu}}$ C
scattering events.
The smooth curve is a gaussian fit with parameters P1-P3 plus a constant (P4).}
\end{figure}

\subsubsection*{6.3.3\ \ Angle Accuracy}%

The Michel electron sample was fit by the reconstruction algorithm described in
section
 6.2.
In Fig. 22 is shown an angular distribution of charge weighted
hit PMTs with respect to the fitted direction of the electron, together
with results from the Monte Carlo simulation.
These electrons have passed the same particle ID criteria as those that appear
in Fig. 20.

\begin{figure}[ht]
\centerline{\psfig{figure=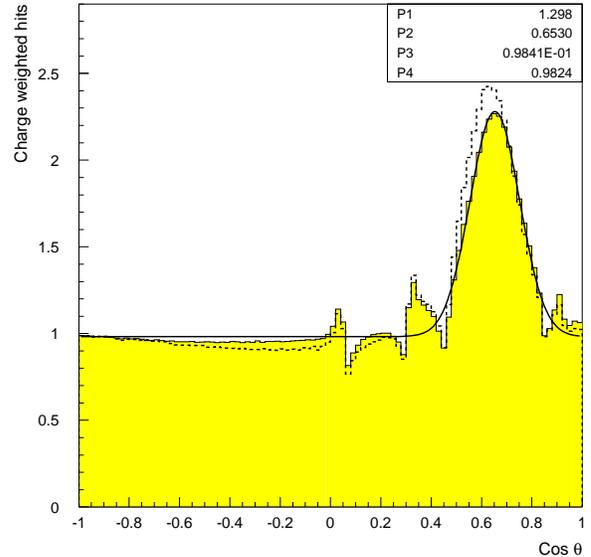,width=3.4in,silent=}}
\caption{Angular distribution of PMTs weighted by pulse height with
respect to the fitted electron direction for muon decay electrons,
data (solid) and Monte Carlo (dashed).
The smooth curve is a fit to a constant (P4) plus gaussian with the parameters
 P1 - P3.}
\end{figure}

This distribution shows a clear peak due to {$\check{\rm C}$erenkov } light at
 cos $\theta \sim$ 0.68 and an approximately isotropic distribution due to
reemitted {$\check{\rm C}$erenkov } light and scintillation light.
The width of the peak is dominated by multiple scattering of electrons in the
detector medium and agreed with the Monte Carlo simulation.
The fine structure away from the {$\check{\rm C}$erenkov } peak is due to the
weighting in position and angle fits.
At 50 MeV, multiple scattering limited angular resolution to about
$12^{\circ}$.
In Fig. 23 is shown the angular distribution for neutrino induced
 electron events in the detector, where $\theta$ is the angle between the
electron direction and the incident neutrino.
The peak in the distribution at cos ${\theta}$ = 1 is from neutrino-electron
scattering and had a width which confirms an angular spread of
$\sim 12^{\circ}$ for electrons above 20 ${\,{\rm~MeV}}$.

\begin{figure}[ht]
\centerline{\psfig{figure=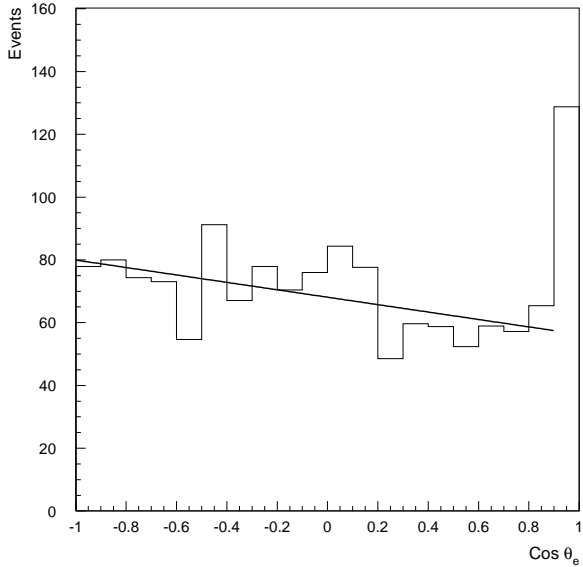,width=3.4in,silent=}}
\caption{Angular distribution for neutrino events with electrons
in the final state with respect to the neutrino beam direction.
The straight line is consistent in slope to that expected from $\nu_e$ C
scattering.}
\label{nue_elastic}
\end{figure}

\subsection*{6.4\ \ Energy Calibration}%

 Muons that stopped in the material of the detector were an excellent
source for energy calibration.
After a small background subtraction of non-correlated events, the charge
distribution was fitted to a Michel decay spectrum convoluted with a
Gaussian energy resolution function that gave the charge-to-energy conversion
factor and the energy resolution at the endpoint.
This Gaussian energy resolution was assumed to vary as $1/\sqrt E$, although
results were not sensitive to the choice of energy dependence.
A resolution of 7.7\% at 52.8 ${\,{\rm~MeV}}$ was derived averaged over the
entire volume with $dist > 25{\,{\rm cm}}$.
This sample was sufficiently large that it was possible to measure the
endpoint energy for bins in a set of parameters discussed below.
A fit was made to the Michel spectra to determine the energy response in each
 bin.
Relative corrections were applied to set the end point of each of these
spectra to 52.8 ${\,{\rm~MeV}}$.
A correction was made for this variation by separating data into bins
according to the value of parameters discussed below.
These corrections were determined and applied sequentially.
Overall run-number dependent correction factors were also applied.
A fully corrected distribution is shown in Fig. 24.
The fit to this distribution gave an energy resolution at the end point of
6.6\%, a 15\% improvement.
In Fig. 25 is shown a comparison of the energy for a sample of
Michel electrons together with a Monte Carlo calculation.

\begin{figure}[ht]
\centerline{\psfig{figure=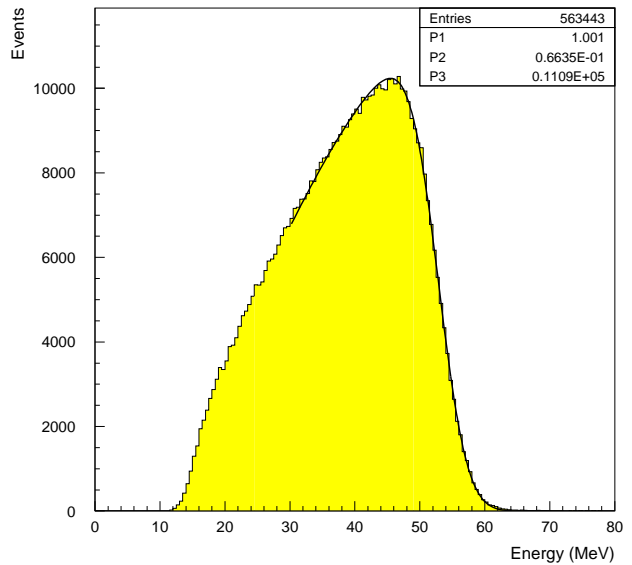,width=3.4in,silent=}}
\caption{Electron energy distribution from stopped muon decay in the medium of
the detector}
\end{figure}

\begin{figure}[ht]
\centerline{\psfig{figure=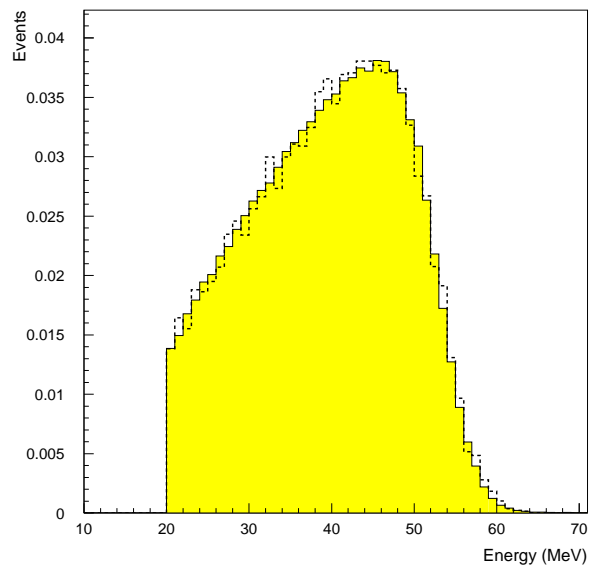,width=3.4in,silent=}}
\caption{Charge distribution for Michel electrons from data (solid) and
simulation (dashed)}
\end{figure}

Part of the calibration adjustment was due to physical processes understood
from Monte Carlo simulation.
In particular, the apparent charge deposited by an electron gradually increased
as its position changed from the center of the tank to $dist \sim 25
{\,{\rm cm}}$, and then dropped sharply in the outer regions of the tank.
There were also variations depending on the direction of the electron with
respect to the center of the tank.
A comparison of charge distributions from Monte Carlo and data Michel events,
as shown in Fig. 25 shows satisfactory agreement.
The distributions agree well in shape, and a gain adjustment of $< 10\%$ was
consistent with expected uncertainties in the amount of scintillation light
(the parameter $A$ in section 5.2) and overall gain factors.
A second portion of the calibration adjustment was due to a periodic
variation of the charge response of the QT cards with the relative phase of
 the event and the 100 ns clock.
The remaining unexplained portion might be due to residual calibration
inaccuracies; a mostly monotonic dependence on the reconstructed z position
 is in this category.

The energy response was corrected in bins in the following parameters:
(1)  The distance from the PMT face to the reconstructed position of the
electron
($\pm 3 \%$);
(2)  The relative phase between PMT hits and the 100 ns clock ($\pm 3.5 \%$);
(3)  The cosine of the angle between the reconstructed electron direction
and the position vector from the center of the tank. ($\pm 7 \%$);
(4)  Reconstructed position along the z axis of the detector($8 \%, 4 \%$);
(5)  The angle of the reconstructed direction with respect to the y axis.
($\pm 1.5 \%$).
The numbers in parentheses are full ranges of typical variation for 1994
and 1995 data respectively.
Because some of the effects were correlated, these ranges may not represent
 the full underlying variations.
Corrections were made iteratively until the remanent corrections were less than
$\pm 0.5 \%$.
The resolution was found to be $\approx 1\%$ larger in the outermost part of
the fiducial volume than in the center of the tank.
In a forthcoming publication \cite{ebigpaper} an analysis of ${\nu_e}$ on
$^{12}C$ scattering to an electron and the ground state of nitrogen,
$^{12}N_{gs}$ will be described.
For this reaction the end point of the electron spectrum is at
35.4${\,{\rm~MeV}}$ and the endpoint of the
$\beta$ decay spectrum from $^{12}N_{gs}$ is at 16${\,{\rm~MeV}}$.
Each of these end points verified the expected linearity of the
detector for electron energy in this range.

\subsection*{6.5\ \ Particle Identification}%

During normal operation, primary events had fewer than four hit PMTs in
the veto in time.
To obtain a sample of cosmic ray neutrons, these events were subsequently
required to have a correlated 2.2-MeV ${\gamma}$ using
criteria described in reference \cite{bigpaper2} and to occur in the beam-off
part of data taking.
{}From the distribution of time between the primary signal and the correlated
${\gamma}$ signal it was determined that the selected sample consisted of
90\% neutrons.
This set of events was used as a ``neutron'' sample to verify the operation
 of the particle identification algorithm.
The ``neutron'' sample was then limited in deposited energy in the detector to
correspond to that for electrons of kinetic energy between 36 and 60
${\,{\rm~MeV}}$.
A sample of electron candidates was selected from the ``Michel'' data set as
described above.
The particle identification algorithm using spatial reconstruction, angular
reconstruction, and late light criteria was applied, and a histogram of the
particle
 ID parameter is shown in Fig. 26 for electrons and for
identified neutrons.

\begin{figure}[ht]
\centerline{\psfig{figure=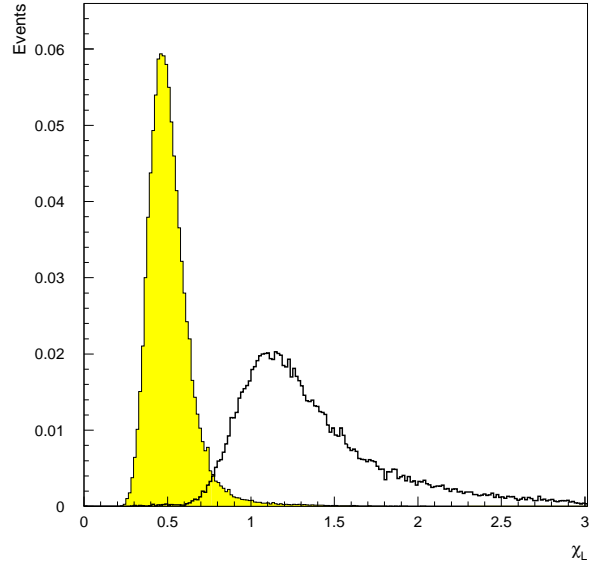,width=3.4in,silent=}}
\caption{Particle ID parameter for `` electrons'' (shaded) and ``neutrons''.}
\end{figure}

These distributions show the difference between electrons and
neutrons as identified by a correlated ${\gamma}$.
Events were selected based on individual $\chi$ parameters and then finally a
cut on the particle ID parameter at $\chi_{tot}$ = 0.65 gave 95\% efficiency
for selecting electrons above $36{\,{\rm~MeV}}$, and a rejection of 10$^{-3}$
against neutrons.
Details of this analysis may be found in \cite{bigpaper2}.

\subsection*{6.6\ \ Duty factor measurements}%

The fraction of the time that the beam was on is referred to as the duty
factor.
The duty factor was subject to variation during the course of
data taking from two basic sources.
The accelerator operated in approximate synchronism with the AC line frequency.
The AC line frequency varied in the course of a
24 hour period so that the time of AC zero crossing could be different
 by as much as a few seconds from uniform time, implying that the accelerator
start time varied also.
Under normal operation, the duration of the beam pulse was constant and
controlled by a precise clock in the accelerator complex.
The effect of this AC variation was to induce a small variation in the duty
factor.
In practical terms however, this variation was negligible.
Accelerator operation was very sensitive to beam loss, and whenever significant
 loss was detected, the ion source current was interrupted and acceleration was
terminated.
This caused significant variation in the length of the beam pulse and, hence, a
short term variation in the duty factor.
Under tuning of the accelerator for minimum beam loss, this condition
occasionally persisted for periods that caused the duty factor to vary
significantly.
An H$^+$ signal was asserted during beam acceleration and for a short period
both ahead and after the actual accelerating period, as described in section
2.1.2.
The variation of the duty factor from beam duration variations was the largest
 effect and amounted to as much as 10\% over periods of minutes.
The duty factor was determined from the ratio of the number of primary triggers
that occur when the H$^+$ signal was on to all primary triggers and was found
to be $0.065 \pm 0.001$ in 1993 - 1995.
Genuine neutrino interactions represented only $\sim 10^{-5}$ of all triggers
 and were a negligible perturbation of the ratio.
Beam induced events that were not neutrino induced were much less than
neutrino induced events and were also negligible.
A ratio of these counts was made over the entire data taking period.
This ratio was then used to multiply the event rates that occurred during
beam-off time to estimate a corresponding rate for beam-on and, hence, provide
a sample for background subtraction.
Although the exact duty factor should be calculated as a function of time
and an average taken, it was determined that the ratio of counts for H$^+$ on
to off, each separately averaged over the data taking period, was the same for
first order variations of the instantaneous duty factor and of the gain of
the detector.

The duty factor was also determined for ${\gamma}$ events (PMT hits $<100$),
 $\beta$ events ($100 < $PMT hits $<250$), and higher energy events
(PMT hits $> 250$) and found to be the same within errors.
Knowing the duty factor, a beam-on minus beam-off subtraction was performed
 to remove the background from beam-unrelated processes.
For a duty factor of $0.065 \pm 0.001$, the ratio of beam-on
events to beam-off events was $0.070 \pm 0.001$.
Therefore, for any event sample the net number of beam-related events is the
number of beam-on events minus 0.070 times the number of beam-off events.

\section* {7. \ \ Conclusion}%

In conclusion, the construction of the LAMPF neutrino source has been
described, including calculations of neutrino flux from the each of the
relevant reactions from pions and muons for each of the three years of
operation.
The expectation that this source will have a low contamination from
$\bar\nu_{e}$ has been justified.
The construction and operation of the detector has been discussed with
a description of salient operating parameters.
The detector has performed well and has been well suited for the measurement
of low energy neutrino interactions with both electrons and muons as the
final state lepton.
Future publications will discuss the search for neutrino oscillations and the
measurement of neutrino-carbon scattering.

%

\pretolerance=10000 \tolerance=10000 \hyphenpenalty=10000
\hbadness=10000 \vbadness=10000
%

%

\paragraph*{Acknowledgements}

This detector system has relied on the efforts and support of many people in
addition to the authors.
Camilo Espinoza, Greg Hart and Neil Thompson have been crucial to the
construction and maintenance of the system.
Butch Daniels, Bill Marterer, and Suzy Weaver were also central to the
construction effort.
Mike Sullivan contributed in the early stages of the experiment.
Conventional construction was important to this experiment and Bill Wassmund
as construction manager proved invaluable.
Earl Hoffman and the accelerator operations personnel produced beam of
excellent quality throughout the data taking period.
The beam stop and neutrino source were demanding to operate at the radiation
levels that were routinely encountered, and the efforts of Dick Werbeck and
his group are truly appreciated.
We were extremely fortunate to have had many undergraduate students who made
 substantial contributions to the experiment; they were acknowledged by
name in a previous letter.
This experiment was built and carried out in a difficult period for this
facility; without the support of Peter Barnes, Cyrus Hoffmann and John
McClelland this work would not have been completed.
This work was directly supported by the U. S. Department of Energy and by
the National Science Foundation.



\begin{references}
\bibitem{ludox}
Ludox was made by Dupont Chemicals, East Chicago, Indiana.
It was a colloidal suspension of 22nm quartz spheres with
a mean free path of about 1.6 {\,{\rm cm}}.
\bibitem{coda}
CEBAF On-line Data Acquisition.  This code was developed
at CEBAF and an early version was tested and debugged in this installation
by this collaboration.
\end{references}
\end{document}